\documentclass[a4paper,11pt]{article}
\pdfoutput=1 

\usepackage{jheppub} 

\reversemarginpar

\usepackage{graphicx}
\usepackage{amsmath}
\usepackage{hyperref}
\usepackage{pdfpages}
\usepackage[section]{placeins}
\usepackage{epstopdf}

\usepackage{tensor}
\usepackage[colorinlistoftodos]{todonotes}
\usepackage{epsfig}
\usepackage{graphicx,color}
\usepackage{cancel}
\usepackage[utf8]{inputenc}
\usepackage{simpler-wick}
\usepackage{comment}

\newcommand{\eps}{\varepsilon}

\newcommand{\nn}{\nonumber}

\newcommand{\veir}{\varepsilon_{\rm {IR}}}
\newcommand{\veuv}{\varepsilon_{\rm {UV}}}

\newcommand{\be}{\begin{equation}}
\newcommand{\ee}{\end{equation}}
\newcommand{\bea}{\begin{eqnarray}}
\newcommand{\eea}{\end{eqnarray}}
\newcommand{\balign}{\begin{align}}
\newcommand{\ealign}{\end{align}}
\newcommand{\as}{\alpha_s}

\newcommand{\bra}[1]{\left< #1 \right |}
\newcommand{\ket}[1]{\left | #1 \right >}
\newcommand{\braket}[1]{\left< #1 \right >}
\newcommand{\sandwich}[3]{\left< #1 \right | #2 \left | #3 \right >}

\newcommand{\bg}{\begin{gather}}
\newcommand{\foma}{\end{gather}}

\newcommand{\noopsort}[1]{}

\def\ve{\varepsilon}

\def\pd{\partial}

\def\<{\langle}
\def\>{\rangle}

\def\d{\delta}

\def\m{\mu}

\def\({\left(}
\def\[{\left[}
\def\){\right)}
\def\]{\right]}

\def\ln{\hbox{ln}}

\def\Slash#1{{#1\!\!\!\slash}}

\newcommand{\ben}{\begin{eqnarray}}
\newcommand{\een}{\end{eqnarray}}

\newcommand{\bef}{\begin{figure}[htb]\centering}
\newcommand{\eef}{\end{figure}}

\newcommand{\textoverline}[1]{$\overline{\mbox{#1}}$}

\usepackage[normalem]{ulem} 

\newcommand{\Q}[4]{ {}^{#1} #2 ^{[#4]}_{#3} }

\newcommand{\NOTE}[1]{\marginpar{\footnotesize\textbf{\color[rgb]{0.9,0,0.9}**NOTE**}}{\color[rgb]{0.9,0,0.9}\sf [#1]}}

\title{Gluon TMD fragmentation function into quarkonium}
\author[a,b]{Miguel G. Echevarria,}
\author[a,b]{Samuel F. Romera}
\author[c]{and Ignazio Scimemi}
\affiliation[a]{Department of Physics, University of the Basque Country UPV/EHU,\\ PO Box 644, 48080 Bilbao, Spain}
\affiliation[b]{EHU Quantum Center, University of the Basque Country UPV/EHU}
\affiliation[c]{Dpto. de Física Teórica $\&$ IPARCOS, Universidad Complutense de Madrid, 28040 Madrid, Spain}

\emailAdd{miguel.garciae@ehu.eus}
\emailAdd{samuel.fernandez@ehu.eus}
\emailAdd{ignazios@ucm.es}

\date{\today}

\abstract{
We compute the gluon transverse-momentum-dependent fragmentation function (TMDFF) at next-to-leading order (NLO) into heavy quarkonium in the color-octet $^3S_1^{[8]}$ channel, based on the NRQCD factorization approach.
The spurious rapidity divergences are explicitly shown to cancel in a well-defined TMDFF, which incorporates the needed soft factor.
We also compute the integrated gluon FF at NLO in the same $^3S_1^{[8]}$ channel, and show that the matching coefficient of the TMDFF onto the FF at large transverse momentum is the expected one.
These results are relevant to perform precise and sensible phenomenological studies of transverse-momentum spectra of quarkonium production, for which the production mechanism through fragmentation plays a relevant role, like in the future Electron-Ion Collider.
}

\preprint{IPARCOS-UCM-044}
\begin{document}
\maketitle


\section{Introduction}

Quarkonium production has recently gained quite some attention as a tool to probe nucleon multi-dimensional structure, in particular transverse-momentum-dependent distributions (TMDs), see e.g. \cite{Angeles-Martinez:2015sea,Chapon:2020heu,Arbuzov:2020cqg}.
In particular, different processes have been proposed as a probe of gluon TMD parton distribution functions (TMDPDFs) \cite{Echevarria:2015uaa,Mulders:2000sh} in both hadron-hadron and lepton-hadron colliders (see e.g.~\cite{Boer:2012bt,Ma:2012hh,Zhang:2014vmh,Boer:2015uqa,Bain:2016rrv,Mukherjee:2015smo,Mukherjee:2016cjw,Lansberg:2017tlc,Bacchetta:2018ivt,DAlesio:2019qpk,Echevarria:2019ynx,Fleming:2019pzj,Scarpa:2019fol,Grewal:2020hoc,Boer:2020bbd,
DAlesio:2020eqo,Boer:2021ehu,DAlesio:2021yws,Kishore:2022ddb,Bor:2022fga,Chakrabarti:2022rjr,Boer:2023zit}).
In these studies, in general, the approach to quarkonium production relies on effective field theories (EFTs), such as the non-relativistic QCD (NRQCD)~\cite{Bodwin:1994jh}, models like the color evaporation model~\cite{Fritzsch:1977ay, HALZEN1977105, Gluck:1977zm}, or just factorization ansatzs.

In more recent years, with the formulation of soft collinear effective theory (SCET) \cite{Bauer:2000ew,Bauer:2001yt,Bauer:2001ct,Beneke:2002ph} new details have been included in this description, providing a robust TMD factorization theorem for the production of quarkonium at small transverse momentum right from the hard reaction, which is given in terms of the so-called and newly introduced TMD shape functions~\cite{Echevarria:2019ynx,Fleming:2019pzj}. 

Despite the abundant interest in quarkonium TMDs, little has been done so far in the direction of TMD quarkonium fragmentation processes, by which we mean single parton fragmentation mechanism.
In processes like $ep\rightarrow J/\psi+ \text{jet}$, $e^+e^-\rightarrow J/\psi +\text{jet}$ or $e^+e^-\rightarrow (J/\psi) (J/\psi)$ and even more energetic ones where the $J/\psi$ is traded for an $\Upsilon$, the quarkonium states can also be produced via the fragmentation of a single parton.
This mechanism becomes relevant when a hard scale, much larger than the quarkonium mass, exists.
Such processes can certainly be observed at an Electron Ion Collider (EIC) as the ones expected to be built in several locations~\cite{AbdulKhalek:2021gbh,Anderle:2021wcy}.

While the light-quark TMD fragmentation function to quarkonia has been considered in \cite{Echevarria:2020qjk} (see also the recent works \cite{Copeland:2023wbu,Copeland:2023qed} for polarized TMDFFs), in this work we concentrate on the gluon TMD fragmentation function to quarkonia.

In our analysis we employ the NRQCD factorization conjecture, where the quarkonium state is produced at large distances through the hadronization of a heavy quark-antiquark pair, $Q\bar{Q}(n)$. 
The pair can be found in any color and angular configuration $n=\Q{2S+1}{L}{J}{\text{col.}}$, but then the probability that the pair decays in the colorless quarkonium state scales with the relative velocity, $v$, of the quark-antiquark pair in the quarkonium rest frame.
We decompose the TMDFF within NRQCD in terms of calculable short-distance matching coefficients and long-distance matrix elements (LDMEs). 
We proceed then to the NLO calculation of this function, extracting the matching coefficient onto the corresponding LDMEs in the region $q_T \sim M$. 
Using this information we can evaluate the contribution to the cross section from gluon fragmentation.
We also compute the matching of the TMDFF onto the corresponding integrated FF at $q_T>M$.

The so-called short-distance coefficient in NRQCD for a gluon fragmenting into a heavy-quark pair in the $^3S_1^{[8]}$ channel has been calculated up to next-to-leading order (NLO) in the strong coupling in \cite{Ma:1995ci,Braaten:2000pc,Ma:2013yla}, with some discrepancies in the finite terms.
We have checked this result as a by-product of our analysis, and we do agree with the result obtained in \cite{Ma:2013yla}.

We regularize the rapidity divergences using a $\delta$-regulator whenever necessary (see~\cite{Echevarria:2016scs,Vladimirov:2021hdn} for more details) and we check explicitly the cancellation of the rapidity divergences with the needed soft factor, which is the same as the one appearing e.g. in Higgs boson production at small transverse momentum~\cite{Echevarria:2015uaa,Echevarria:2015byo}.

This paper is organized as follows.
In Sec.~\ref{sec:Notation} we start by setting the notation and the relevant definitions.
Then in Sec.~\ref{sec:Results} we present the main results, for which more technical details can be found in the Appendix. Finally in Sec.~\ref{sec:Conclusions} we conclude.

\section{Notation}
\label{sec:Notation}
In this Section we establish the notation that we are going to use in the calculations, as well as the general definition of the gluon TMD fragmentation function and its matching onto the LDMEs.

\subsection{Kinematics}

We use the coordinates of the light cone set by the following scalar products in the space-time dimensions $d=4-2\varepsilon$:
\begin{gather}
    p^2 = 2 p^+ p^- - p_T^2,\\
    q \cdot p = q^+ p^- + q^- p^+ + \mathbf{q}_\perp \cdot \mathbf{p}_\perp, \\
    g_T^{\mu \nu} = g^{\mu \nu} - n^\mu \bar{n}^\nu - \bar{n}^\mu n^\nu ,
\end{gather}
with $n^2 = \bar{n}^2 =0$, $(\bar{n} n)=1$. 
We denote the momentum of the fragmenting gluon by $q$ and the momentum of the heavy-quark $Q \bar{Q}$ pair by $P$.
Also, we work in the frame in which the transverse momentum of $P$ vanishes.

\subsection{Gluon TMDFF definition}
The operator for the gluon fragmentation function follows from \cite{Echevarria:2016scs}.
The unsubtracted TMDFF is the hadronix matrix element of that operator
\begin{gather}
\label{eq:TMDFFnorenor}
    \Delta_{g \rightarrow J/\psi} (z, \mathbf{b}_T ) = \frac{-P^+}{2(1- \varepsilon) (N_c^2 -1)} \sum_X \int \frac{d \xi^-}{2 \pi} e^{-i P^+ \xi^-/z}\\
    \times \bra{0}T \left[ \mathcal{B}_{n \perp}^\mu \right] \left(  \frac{\xi}{2} \right)  \ket{X, J/\psi} \bra{X, J/\psi} \bar{T} \left[ \mathcal{B}_{n \perp \mu} \right] \left( \frac{-\xi}{2} \right)   \ket{0} ,\nonumber
\end{gather}
where $\xi=\{0^+,\xi^-,\mathbf{b}_T\}$ and $\mathbf{b}_T$ is the conjugate variable to the transverse momentum. 
The color normalization factor and the normalization of the number of physical gluon polarizations in $d = 4 - 2 \varepsilon$ are encoded in the prefactor. In this equation, $\mathcal{B}_{n \perp}^\mu$ is the gluon field strength defined as follows
\begin{equation}
\begin{aligned} \label{eq:Bfield}
\mathcal{B}_{n \perp}^\mu & = \frac{1}{g_s} \left[ W_n^\dagger (y) \, i D_{n \perp}^\mu \, W_n(y) \right] ,
\end{aligned}
\end{equation}
with $i D_{n \perp}^\mu = \partial_{n \perp}^\mu + g_s A_{n \perp}^\mu$ where $A_{n \perp}^\mu$ is the SCET $n$-collinear field.
In (\ref{eq:Bfield}), $W_n$ is the collinear Wilson lines defined as
\begin{gather}
W_n(y) = P \exp \left[ i g_s \int_{- \infty}^0 ds \, \bar n \cdot A_n (y + \bar n s) \right],
\\ \nonumber  
W_n^\dagger (y) = P \exp \left[ - i g_s \int_{- \infty}^0 ds \, \bar n \cdot A_n (y + \bar n s) \right]
\,.
\end{gather}

The renormalized gluon TMD is defined as follows
\begin{gather}
    \label{eq:renormalizedTMDFF}
    D_{g \rightarrow J/\psi} (z, \mathbf{b}_\perp, \mu, \zeta ) = Z_g (\mu, \zeta) R_g (\mu, \zeta) \Delta_{g \rightarrow J/\psi} (z, \mathbf{b}_\perp) \, ,
\end{gather}
where $Z_g$ is the usual renormalization factor for UV divergences and $R_g$ is the rapidity renormalization factor, $\mu$ is the scale of UV subtraction and $\zeta$ is the scale of rapidity subtraction. Here, $R_g$ is the following expression
\begin{gather}
    \label{eq:Rg}
    R_g (\mu, \zeta) = \frac{\sqrt{S(\mathbf{b}_\perp)}}{\mathbf{Z_b}},
\end{gather}
which describes the ratio between the soft function denoted as $S({\mathbf{b}_\perp})$ and the soft overlap of the collinear and soft sectors through the term $\mathbf{Z_b}$ \cite{Manohar:2006nz},  denoting the zero-bin contribution.

The soft function is defined as a expectation value of soft Wilson lines:
\begin{gather} \label{eq:fullSF}
S(\mathbf{b}_\perp) = 
\frac{1}{N_c^2-1}
\sum_{X_s}
\sandwich{0}
{\Big({\cal S}_n^{ \dagger} \tilde{\cal S}_{\bar{n}}\Big)^{ab} (0^+,0^-,\mathbf{b}_\perp)}{X_s}
\sandwich{X_s}
{\Big(
\tilde{\cal S}_{\bar{n}}^{ \dagger} 
{\cal S}_n
\Big)^{ba}(0)}
{0} ,
\end{gather}
where the soft Wilson lines are defined as
\begin{gather}
{\cal S}_n(x) = 
P \exp \left[ i g_s \int_{- \infty}^0 ds \, n \cdot A (x+sn) \right]
, \\
\tilde{{\cal S}}_{\bar{n}}(x) = P \exp \left[- i g_s \int_{- \infty}^0 ds \, \bar{n} \cdot A (x+s \bar n) \right]. \nonumber
\end{gather}
These soft Wilson lines, with calligraphic typography, are in the adjoint representation, where the color generators are given by $(t^a)^{bc} = -i f^{abc}$.

\subsection{Gluon TMDFF factorization}

We employ, for $q_T \sim M$, the NRQCD formalism \cite{Bodwin:1994jh} to write the gluon TMDFF as a product of short distance coefficients and the long distance matrix elements (LDMEs):
\begin{gather}
    \label{eq:TMDFFfactorization}    
    D_{g \rightarrow J/\psi} (z, \mathbf{b}_\perp) = \sum_{n} d_{g \rightarrow Q \bar Q (n)}(z, \mathbf{b}_\perp) \braket{\mathcal{O}_{n}^{J/\psi}} ,
\end{gather}
Here, $n = \tensor[^{2S+1}]{L}{_J^{[col.]}}$ describes the color and angular momentum configuration of the heavy-quark pair. All relativistic effects are absorbed in $d_{g \rightarrow Q \bar Q (n)}$, which can be calculated as a perturbative series in the strong coupling constant $\alpha_s$ through matching, and the LDMEs are defined as follows
\begin{gather}
    \label{eq:LDMEs}
    \braket{\mathcal{O}_{n}^{J/\psi}} = \bra{0}\chi^\dagger \mathcal{K}_n \psi a_{J/\psi}^\dagger a_{J/\psi} \psi^\dagger \mathcal{K}_n' \chi \ket{0} ,
\end{gather}
where $a_{J/\psi}$ and $a_{J/\psi}^\dagger$ are the operators of annihilation and creation of the state describing the $J/\psi$, $\mathcal{K}_n$ and $\mathcal{K}_n'$ are products of a color matrix, a spin matrix and other fields, and $\chi$ and $\psi$ are the field operators for the heavy quarks in NRQCD.


\section{Results}
\label{sec:Results}

In the present section we show the results of the calculation of the gluon TMDFF for quarkonium at NLO, defined in \eqref{eq:renormalizedTMDFF}.
The details of the calculation can be found in the Appendix.
We have used the $\delta$-regularization \cite{Echevarria:2015byo} in order to regularize the rapidity divergences, defined at operator level as follows:
\begin{equation}
\begin{aligned}
W_n(y) & \longrightarrow  P \exp \left[ i g_s \int_{- \infty}^0 ds \, \bar n \cdot A_n (y + \bar n s) \, e^{-\delta^+ s} \right]
\,,
\end{aligned}
\end{equation}
and similarly for the rest of the collinear and soft Wilson lines.




\subsection{Leading Order}

On the one hand, we calculate the left hand side of the equation \eqref{eq:TMDFFfactorization}, i.e. the matrix element \eqref{eq:TMDFFnorenor}, around the threshold ($\mathbf{q} = 0$):
\begin{equation}
\begin{aligned}
\Delta^{\text{LO}}_{g \rightarrow J/\psi} & = \frac{\delta(1-z)}{8 (d-2)} \frac{g_s^2 (d-2)}{16 m_c^4} \frac{4 m_c^2}{(d-1)} \, \xi^\dagger \, \sigma^k T^a \, \eta \times \eta^\dagger \, \sigma^k T^a \, \xi \\
& = \frac{\pi \alpha_s}{8(d-1)m_c^2} \delta(1-z) \xi^\dagger \, \sigma^k T^a \, \eta \times \eta^\dagger \, \sigma^k T^a \, \xi .
\end{aligned}
\end{equation}
This result is already the TMDFF, since the soft function at LO is just 1.
The spinorial structure which we obtain describes the configuration $n = \tensor[^3]{S}{_1^{[8]}} $.
\par On the other hand, we know from pNRQCD that the LDME describing this configuration is the following,
\begin{gather}
    \braket{\mathcal{O}^{J/\psi}(^3 S_1^{[8]})} \left.  \right|_{pNRQCD} = m_c \, \xi^\dagger \, \sigma^k T^a \, \eta \times \eta^\dagger \, \sigma^k T^a \, \xi .
\end{gather}
Therefore, by matching both sides of the equation \eqref{eq:TMDFFfactorization} the SDC is
\begin{gather}
\label{eq:LO}
d_{g \rightarrow J/\psi}^{\text{LO}}(z,\mathbf{b}_\perp) = \frac{\pi \alpha_s}{8(d-1) m_c^3} \delta(1-z) 
\,.
\end{gather}

\subsection{Next to Leading Order}

The virtual contributions to the gluon TMDFF at NLO are shown in figure \ref{fig:NLOvirtual}.
\begin{equation}
\begin{aligned} \label{eq:VirNorenor}
d^{\text{NLO,vir.}}_{g \rightarrow J/\psi} (z ; \delta) & = \frac{\pi \alpha_s}{8 (d-1) \, m_c^3 } \frac{\alpha_s C_A}{2 \pi} \delta(1-z) \left[ \frac{1}{\veuv} \left(\frac{\beta_0}{ C_A} + 2\, \ln \delta \right) - \frac{1}{\veir }  \right. \\
 & -  \left. 2 \, \ln^2 \delta + 2\, \ln \delta \, \ln \frac{\mu^2}{M^2} +  \left( \frac{8}{3} - \frac{2 n_f}{3 C_A} \right) \ln \frac{\mu^2}{M^2} + \frac{16}{3} \ln{2} - \frac{3\pi^2}{2} + \frac{59}{9} - \frac{10 n_f}{9 C_A}  \right] ,
\end{aligned}
\end{equation}
where $\delta \equiv \delta^+/P^+$ and $\beta_0 = 11 C_A/3-2 n_f/3$. 
In order to obtain a well-defined hadronic quantity it is necessary to renormalize the divergences according to the equations \eqref{eq:renormalizedTMDFF} and \eqref{eq:Rg}. 
We have used the $\delta$-regularization where the subtractions related with $\mathbf{Z_b}$ are equal to the soft function \cite{Echevarria:2016scs}:
\begin{gather}
    \label{eq:deltaRg}
    R_g(\mu,\zeta) = \frac{1}{\sqrt{S(\mathbf{b}_\perp; \mu,\zeta)}}\, .
\end{gather}
The virtual contribution to the SF at one loop is as follows
\cite{Echevarria:2015uaa,Echevarria:2015byo}
\begin{equation}
\begin{aligned} 
S^{\rm{vir.}} ( \delta^+ ,\zeta) & = 
\frac{\alpha_s C_A}{2 \pi}
\left[ 
\frac{-2}{\veuv^2} 
+ \frac{2}{\veuv} 
\ln\frac{\delta^{+2}\zeta}{(P^+)^2\mu^2} 
- 
\ln^2\frac{(\delta^+)^2}{\mu^2}
- \frac{\pi^2}{2} \right] .
\end{aligned}
\end{equation}

At the end, after the renormalization of the rapidity divergences, the virtual contribution of the SDC at NLO is
\begin{equation}
\begin{aligned} \label{eq:NLOvir}
d^{\text{NLO,vir.}}_{g \rightarrow J/\psi} (z; \delta, \mu, \zeta)  = & \, \frac{\pi \alpha_s}{8 (d-1)\, m_c^3}  \frac{\alpha_s C_A}{2 \pi} \delta(1-z) \left[ \frac{1}{ \veuv^2} + \frac{1}{\veuv} \left( \frac{\beta_0}{ C_A} + \ln{\frac{\mu^2}{\zeta}} \right)  \right. \\
& \left.
- \frac{1}{\veir}
+ \frac{1}{2} \ln^2 \frac{\delta^{+2}}{\mu^2} - 2 \ln^2 \frac{\delta^+}{P^+}  +2 \ln{\frac{\delta^+}{P^+}} \ln{\frac{\mu^2}{M^2}} + \left( \frac{8}{3} - \frac{2 n_f}{3 C_A} \right) \ln \frac{\mu^2}{M^2} \right. \\
& + \frac{16}{3} \ln 2 - \left. \frac{5 \pi^2}{4}  + \frac{59}{9} - \frac{10 n_f}{9 C_A}  \right] . 
\end{aligned}
\end{equation}

Since real diagrams will not contain
UV divergences, because the transverse momentum (or distance) is finite, we can extract the evolution of the TMDFF solely from its virtual part.
The renormalization of the TMDFF at NLO requires the renormalization of both $\alpha_s$, which appears already at LO, and the operator itself. 
In the \textoverline{MS}-scheme it is well-known that the coupling is renormalized as follows:
\begin{equation}
\begin{aligned}
    \alpha_s  \longrightarrow & \,  \alpha_s \left(  1  - \frac{\alpha_s}{\pi} \frac{\beta_0}{4} \frac{\left( 4 \pi e^{- \gamma_E}\right)^\ve}{\veuv} \right) \\
    & = \alpha_s \left[ 1 - \frac{\alpha_s}{\pi} \frac{\beta_0}{4} \left( \frac{1}{\veuv} + \ln \frac{4 \pi}{e^{\gamma_E}} + \mathcal{O}(\epsilon) \right) \right] 
\end{aligned}
\end{equation}
And combining this result with the TMDFF at LO in \eqref{eq:LO} and the NLO result in 
\eqref{eq:NLOvir} we obtain the complete renormalization factor for the UV divergences:
\begin{equation}
\begin{aligned}
Z_g (\mu, \zeta) & = 1 - \frac{\alpha_s C_A}{2 \pi} \left[\frac{1}{\veuv^2} + \frac{1}{\veuv} \left( \frac{\beta_0}{2C_A} + \ln \frac{\mu^2}{\zeta} \right) \right].
\end{aligned}
\end{equation}
The anomalous dimension of the TMDFF \cite{Scimemi:2019cmh}, which gives the evolution in the renormalization scale $\mu$ to $\mathcal{O}(\alpha_s)$, is the following
\begin{equation}
\begin{aligned} \label{eq:AnomalousD}
\gamma_{F} (\mu,\zeta)  
&\equiv
\frac{d}{d\ln\mu} \ln Z_g(\mu,\zeta)
= \frac{1}{Z_g (\mu,\zeta)}\frac{\pd Z_g(\mu, \zeta)}{\pd\ln\m}
+ \frac{1}{Z_g(\mu,\zeta)}\frac{\pd Z_g(\mu,\zeta)}{\pd\as}
\Big(-2\eps\as + {\cal O}(\as^2) \Big)
\nn\\
&= \frac{\alpha_s C_A}{2\pi} 
\left( 
\frac{\beta_0}{C_A} 
+ 2\ln \frac{\mu^2}{\zeta}
\right).
\end{aligned}
\end{equation}
This result obviously agrees with the one that can be found e.g. in \cite{Echevarria:2016scs}, since the UV behavior of the operator does not depend on the nature of the hadronic state.\\

We now turn to the real contribution to the gluon TMDFF at NLO, coming from the diagrams shown in figure \ref{fig:NLOreal}. 
We separate the diagrams into two groups. 
Those with rapidity divergences, i.e. $c$, $d1$ and $d2$, and those without, i.e. $a$, $b$, $e1$, $e2$, $f1$ and $f2$. \par

The contribution of diagrams $c$, $d1$ and $d2$ is
\begin{equation}
\begin{aligned}
\label{eq:cdnorenor}
d^{\rm{c,d}}_{g \rightarrow J/\psi} (z, b_T; \delta, \mu) = & \frac{C_A \alpha_s^2}{8(d-1) \, m_c^3} \left[ \delta(1-z) \left( \ln \, \delta \left( L_T - \ln \frac{\mu^2}{M^2} \right)  + \ln^2 \delta + \frac{\pi^2}{12} \right) \right. \\
&  + \left. \frac{-4 z^4+11 z^3-20 z^2+13 z-8}{8 z (1-z)_+} \left( L_T - \ln \frac{\mu^2}{M^2} \right) \right. \\
& \left. + \frac{4 z \left( z^2 - 1 \right) \, A_1(z,b_T) + 8  \left( z^4 - 3 z^3 + 5 z^2 -3 z +2 \right) \, A_2(z,b_T) }{8 z (1-z)_+}  \right. \\
& \left. - \frac{4 z^4 -11 z^3 + 20 z^2 -13z+8}{4z} \left( \frac{\ln(1-z)}{1-z} \right)_+ \right] \, ,
\end{aligned}
\end{equation}
where $L_T = \ln \left( \mu^2 b_T^2 e^{2 \gamma_E}/4 \right)$ and $A_1(z, b_T)$ and $A_2(z, b_T)$ are defined in the Appendix.

The contribution of diagrams $a$, $b$, $e$ and $f$ is
\begin{equation}
\begin{aligned}
\label{eq:abefnorenor}
d^{\rm{a,b,e,f}}_{g \rightarrow J/\psi} (z, b_T; \delta, \mu) & = \frac{C_A \alpha_s^2}{8(d-1) \, m_c^3} \left[ \frac{\delta(1-z)}{2} \left( \frac{1}{\veir} +  \ln \frac{\mu^2}{M^2} \right) \right. \\
& +  \left. \frac{(1-z)(4 z^2 - z + 3)}{8 (1-z)_+} \left( L_T - \ln \frac{\mu^2}{M^2} \right)  \right. \\
& - \left. b_T M  (z^2 - 2 z +2) B(z, b_T) \right. \\
& + \left. \frac{4(1-z^2) A_1(z, b_T) - 8 (1-z)(1+z^2) A_2(z, b_T) }{8(1-z)_+}  \right.  \\
& - \left.  \frac{z (z^2 -2 z + 2)}{(1-z)_+} + \frac{(1-z)(4 z^2 -z+3)}{4} \left( \frac{\ln(1-z)}{1-z} \right)_+  \right] 
\,,
\end{aligned}
\end{equation}
where $B(z, b_T)$ is defined in the Appendix.\\

Adding the results \eqref{eq:cdnorenor} and \eqref{eq:abefnorenor} we obtain the real contribution before renormalizing the rapidity divergences:
\begin{equation}
\begin{aligned} d^{\rm{NLO,real}}_{g \rightarrow J/\psi} (z, b_T ; \delta, \mu) & = \frac{\pi \alpha_s}{8(d-1) \, m_c^3 } \frac{\alpha_s C_A}{2 \pi} \left\{ \delta(1-z) \left( \frac{1}{ \veir } + 2 \ln \, \delta  \left( L_T - \ln \frac{\mu^2}{M^2} \right) + 2 \ln^2 \delta \right) \right. \\
& + \left.  \delta(1-z) \left(  \ln \frac{\mu^2}{M^2} + \frac{\pi^2}{6}  \right) - 2 b_T M (z^2 -2 z +2) B(z,b_T)  \right. \\
& - \left.  P_{g/g} \left[ \left( L_T - \ln \frac{\mu^2}{M^2} \right) - 2  A_2 (z, b_T)\right] \right. \\
& - \left. \frac{2 z(z^2-2z+2)}{(1-z)_+}  - \frac{4 (z^2 -z +1)^2}{z} \left( \frac{\ln(1-z)}{1-z} \right)_+ \right\}
\,,
\end{aligned}
\end{equation}
where
\begin{equation}
\begin{aligned}
P_{g/g} = 2 \left[ \frac{z}{(1-z)_+} + \frac{1-z}{z} + z(1-z) \right]
\,.
\end{aligned}
\end{equation}

The real contribution of the Soft Function, which can be found e.g. in \cite{Echevarria:2015uaa,Echevarria:2015byo}, is
\begin{gather}
    S^{\rm{real}} (b_T; \delta^+, \zeta)  = \frac{\alpha_s C_A}{2 \pi} \left( L_T^2 + \ln^2 \frac{(\delta^+)^2}{\mu^2} + 2 L_T \ln \frac{(\delta^+)^2 \zeta}{(P^+)^2 \mu^2} + \frac{2 \pi^2}{3} \right) \, .
\end{gather}

Therefore, the real contribution to the SDC at NLO after the renormalization of the rapidity divergences is the following:
\begin{equation}
\begin{aligned} \label{eq:NLOreal} 
d^{\rm{NLO, real}}_{g \rightarrow J/\psi} (z, b_T ; \delta^+, \mu, \zeta)  & =   \frac{\pi \alpha_s}{8 (d-1) m_c^3 } \frac{\alpha_s C_A}{2 \pi} \left[ \delta(1-z) \left( \frac{1}{ \veir} - 2 \ln \frac{\delta^+}{P^+} \ln \frac{\mu^2}{M^2}  + 2 \ln^2 \frac{\delta^+}{P^+} \right)  \right. \\
& \left. + \delta(1-z) \left( - \frac{L_T^2}{2}  + L_T \, \ln \frac{\mu^2}{\zeta}- \frac{1}{2} \ln^2 \frac{(\delta^+)^2}{\mu^2} +  \ln \frac{\mu^2}{M^2} - \frac{\pi^2}{6}  \right) \right.  \\
& \left. - P_{g/g} \left[ \left( L_T - \ln \frac{\mu^2}{M^2} \right) - 2 A_2 (z, b_T)\right]  - \frac{2 z(z^2-2z+2)}{(1-z)_+}   \right. \\
& \left. - 2 b_T M (z^2 -2 z +2) B(z,b_T) - \frac{4 (z^2 -z +1)^2}{z} \left( \frac{\ln(1-z)}{1-z} \right)_+ \right]
\end{aligned}
\end{equation}

Finally, by adding the virtual part in \eqref{eq:NLOvir} and the real part in \eqref{eq:NLOreal}, we get the short-distance matching coefficient for the gluon TMDFF at NLO into heavy quarkonium in the color-octet $\tensor[^3]{S}{_1^{[8]}}$ channel:
\begin{equation}
\begin{aligned} \label{eq:NLOtotal}
d_{g \rightarrow J/\psi}^{\text{NLO,bare}} (z, b_T; \mu, \zeta) & = \frac{ \pi \alpha_s}{8(d-1) m_c^3} \, \frac{\alpha_s C_A}{2 \pi} \left[\delta(1-z) \left( \frac{1}{ \veuv^2} + \frac{1}{ \veuv } \left( \frac{\beta_0}{ C_A} + \ln \frac{\mu^2}{\zeta} \right) \right) \right.\\
& \left. + \delta(1-z) \left( - \frac{L_T^2}{2} + L_T \, \ln \frac{\mu^2}{\zeta} - \frac{\pi^2}{12} \right) - P_{g/g} \left( L_T - 2 A_2(z, b_T) \right) \right. \\
& \left. + \delta(1-z) \left(  \beta_0 \, \ln \frac{\mu^2}{M^2} - \frac{4 \pi^2}{3} + \frac{16}{3} \ln 2 + \frac{59}{9}  - \frac{10 n_f}{9 C_A}  \right)   \right. \\
& \left.  + P_{g/g} \ln \frac{\mu^2}{M^2}  - 2 b_T M (z^2 -2 z +2) B(z,b_T)   \right.  \\
& \left.  - \frac{ 2z^3 -4z^2 + 4z}{(1-z)_+} - \frac{4 (z^2 -z +1)^2}{z} \left( \frac{\ln(1-z)}{1-z} \right)_+  \right]
\,.
\end{aligned}
\end{equation}
We have obtained a well-defined quantity, since it is free from rapidity divergences and is well-behaved in the limit $z\to 1$ (in terms of distributions), remaining only the UV divergences which are removed by standard renormalization.\\

We end this section by focusing on the $\zeta$-evolution of the gluon TMDFF. 
The TMD evolution equation in $\zeta$ \cite{Echevarria:2012pw} is the following:
\begin{equation}
\begin{aligned}
\frac{\partial}{\partial \, \ln \zeta}  \ln\, D^{ren.}_{g \rightarrow J/\psi}(z,b_T; \mu,\zeta ) = -\mathcal{D}_g (b_T; \mu) ,
\end{aligned}
\end{equation}
where $\mathcal{D}_g (b_T; \mu)$ is called the rapidity anomalous dimension (RAD), also called Collins-Soper (CS) kernel. 
Therefore, from our results, the RAD which gives the evolution in the rapidity scale $\zeta$ to $\mathcal{O}(\alpha_s)$ is
\begin{equation}
\begin{aligned} \label{eq:RAD}
    \mathcal{D}_g(b_T;\mu) = \frac{\alpha_s C_A}{2 \pi} L_T (b_T; \mu) ,
\end{aligned}
\end{equation}
with $L_T (b_T; \mu) = \ln (\mu^2 b_T^2 e^{2 \gamma_E}/4)$. 
This result, es expected, is the same as for the TMDPDF \cite{Echevarria:2015uaa}, since being the TMD operator the same, the structure of the rapidity divergences is the same.

Finally, the evolution in the plane $(\mu,\zeta)$ of the gluon TMDFF is
\begin{equation}
\begin{aligned}
D^{ren.}_{g \rightarrow J/\psi} (z,b_T;\mu_f,\zeta_f) = 
\exp \left[ \int_P \left( \gamma_F(\mu,\zeta) \frac{d \mu}{\mu} - \mathcal{D}_g (b_T; \mu) \frac{d \zeta}{\zeta} \right) \right] D^{ren.}_{g \rightarrow J/\psi} (z,b_T; \mu_i,\zeta_i)
\,,
\end{aligned}
\end{equation}
with the UV anomalous dimension $\gamma_F (\mu, \zeta)$ and the RAD $\mathcal{D}_g(b_T; \mu)$ obtained at ${\cal O}(\alpha_s)$ in \eqref{eq:AnomalousD} and \eqref{eq:RAD}, respectively. 
In the integral, $P$ denotes any path connecting the points $(\mu_f, \zeta_f)$ and $(\mu_i,\zeta_i)$.

\subsection{Matching onto integrated FF}

Following the notation of \cite{Echevarria:2016scs}, the small-$b_T$ matching between the gluon TMDFF and its corresponding integrated function is described by the OPE of the gluon TMDFF onto the standard FF:
\begin{equation}
\begin{aligned} \label{eq:OPE}
D_{g \rightarrow J/\psi}^{ren.} (z, b_T ; \mu, \zeta) = \sum_{f'} C_{g/f'} (z,b_T; \mu, \zeta) \otimes \frac{D_{f' \rightarrow J/\psi}^{ren.}(z; \mu)}{z^{2-2\varepsilon}} + \mathcal{O}(b_T M) ,
\end{aligned}
\end{equation}
where $\otimes$ is the Mellin convolution in variable $z$, and both hadronic matrix elements are understood to be renormalized. 
All the dependence on the transverse coordinate $b_T$ and rapidity scale is in the OPE Wilson coefficient.
The integrated FF is defined as
\begin{gather}
\label{eq:FFdef}
D_{g \rightarrow J/\psi} (z;\mu) 
= \frac{-z^{2-2\eps} P^+}{2(1- \varepsilon)  (N_c^2 -1)} \sum_X \int \frac{d \xi^-}{2 \pi} e^{-i P^+ \xi^-/z}\\
    \times \bra{0}T \left[ \mathcal{B}_{n \perp}^\mu \right] \left(  \frac{\xi^-}{2} \right)  \ket{X, J/\psi} \bra{X, J/\psi} \bar{T} \left[ \mathcal{B}_{n \perp \mu} \right] \left( \frac{-\xi^-}{2} \right)   \ket{0} ,\nonumber
\end{gather}
Notice the different prefactor of $z$ as compared to the TMDFF (apart from the obvious difference in the separation of the fields of the operator, which in this case is just in the collinear direction).

In Section~\ref{sec:FF} we have calculated the collinear unpolarized FF in order to obtain the OPE Wilson coefficient of the perturbative expansion of the unpolarized TMDFF at large transverse momentum:
\begin{align} 
\label{eq:FF}
\frac{1}{z^2} \,
d_{g \rightarrow J/\psi}^{\text{NLO,bare}} (z; \mu) & =  \frac{\pi \alpha_s }{8(d-1) \, m_c^3 } \frac{\alpha_s C_A}{2 \pi} \left[ \frac{1}{ \veuv} \left(  \frac{\beta_0}{ C_A} \delta(1-z) + P_{g/g} \right) 
\right.  
\nn\\ 
& \left. 
+   \Big(
P_{g/g} + \beta_0 \, \d(1-z) \Big) 
\ln \frac{\mu^2}{M^2}
\right.
\nn\\
&\left.
+ \delta(1-z) \left( - \frac{4 \pi^2}{3} + \frac{16}{3} \ln 2 + \frac{59}{9} - \frac{10 nf}{9 C_A} \right) \right.  
\nn\\
& \left. -  \frac{2z^3 -4 z^2 +4z }{(1-z)_+} - \frac{4(z^2 - z +1)^2}{z}\left( \frac{\ln(1-z)}{1-z} \right)_+  \right] .
\end{align}
With this result we have extracted the matching coefficient in the $g/g$ channel at NLO:
\begin{equation}
\begin{aligned} \label{eq:Cgg}
\mathcal{C}_{g/g} (z, b_T; \mu , \zeta) & = \delta(1-z) \\
& + \frac{\alpha_s C_A}{2 \pi} \left[ \delta(1-z) \left( - \frac{L_T^2}{2} + L_T \, \ln \frac{\mu^2}{\zeta} - \frac{\pi^2}{12} \right) - P_{g/g} \left( L_T  -  2 \ln \, z \right)  \right] .
\end{aligned}
\end{equation}
Here, notice that we have expanded in $b_T M$ the functions
\begin{equation}
\begin{aligned}
A_2(z, b_T) & = \ln \, z + \mathcal{O}(b_T M), \\
(b_T M) B(z, b_T) & = \mathcal{O}\left((b_T M)^2 \right)
\,,
\end{aligned}
\end{equation}
in the TMDFF in \eqref{eq:NLOtotal}, in order to obtain the Wilson matching coefficient in \eqref{eq:Cgg} via equation \eqref{eq:OPE}.
The obtained coefficient is the same as in~\cite{Echevarria:2016scs}, as expected, since the matching is done at the operator level, regardless of the hadronic state.

\section{Conclusions}
\label{sec:Conclusions}
The calculation of the $^3S_1^{[8]}$ gluon TMD fragmentation function at NLO that we have performed in this paper shows how the TMDFF behaves at the threshold of quarkonium production. 
For this we have combined both TMD formalism and NRQCD factorization at leading power. 
As expected, the rapidity divergences typical of TMD functions are not affected by the threshold, because the light-cone and heavy-mass regimes do not interfere with each other.
This means that one can use all our knowledge about the UV and rapidity evolution kernel also in the case of quarkonia production. 
We recall that the rapidity evolution kernel has been recently extracted from experiment with a N$^4$LL analysis~\cite{Moos:2023yfa}.
The results of this paper provide necessary ingredients to perform phenomenological studies of transverse-momentum spectra of quarkonia production, for which the fragmentation production mechanism is important.

\acknowledgments

This project is supported by the State Agency for Research of the Spanish Ministry of Science and Innovation through the grants PID2019-106080GB-C21, PCI2022-132984, PID2022-136510NB-C31, PID2022-136510NB-C33 and CNS2022-135186, by the Basque Government through the grant IT1628-22, as well as by the European Union Horizon 2020 research and innovation program under grant agreement Num. 824093 (STRONG-2020).

\begin{figure}
    \includegraphics[width=\textwidth]{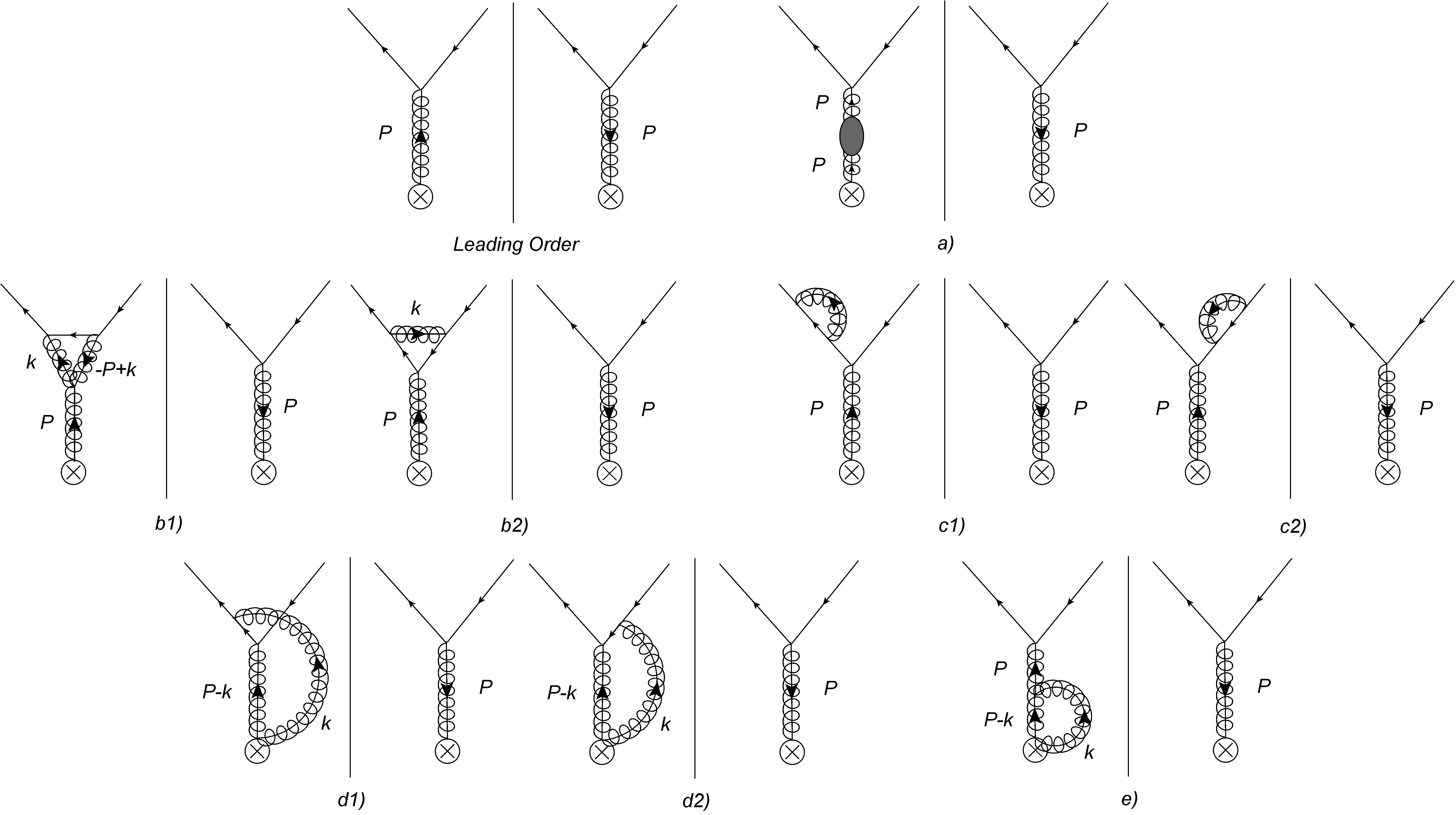}
    \caption{ Diagram contributing to LO and virtual diagrams contributing to NLO. Only diagrams d1, d2 and e have rapidity divergences. Hermitian conjugates of diagrams a, b, c, d1, d2 and e are not shown. }
    \label{fig:NLOvirtual}
\end{figure}

\begin{figure}
    \includegraphics[width=\textwidth]{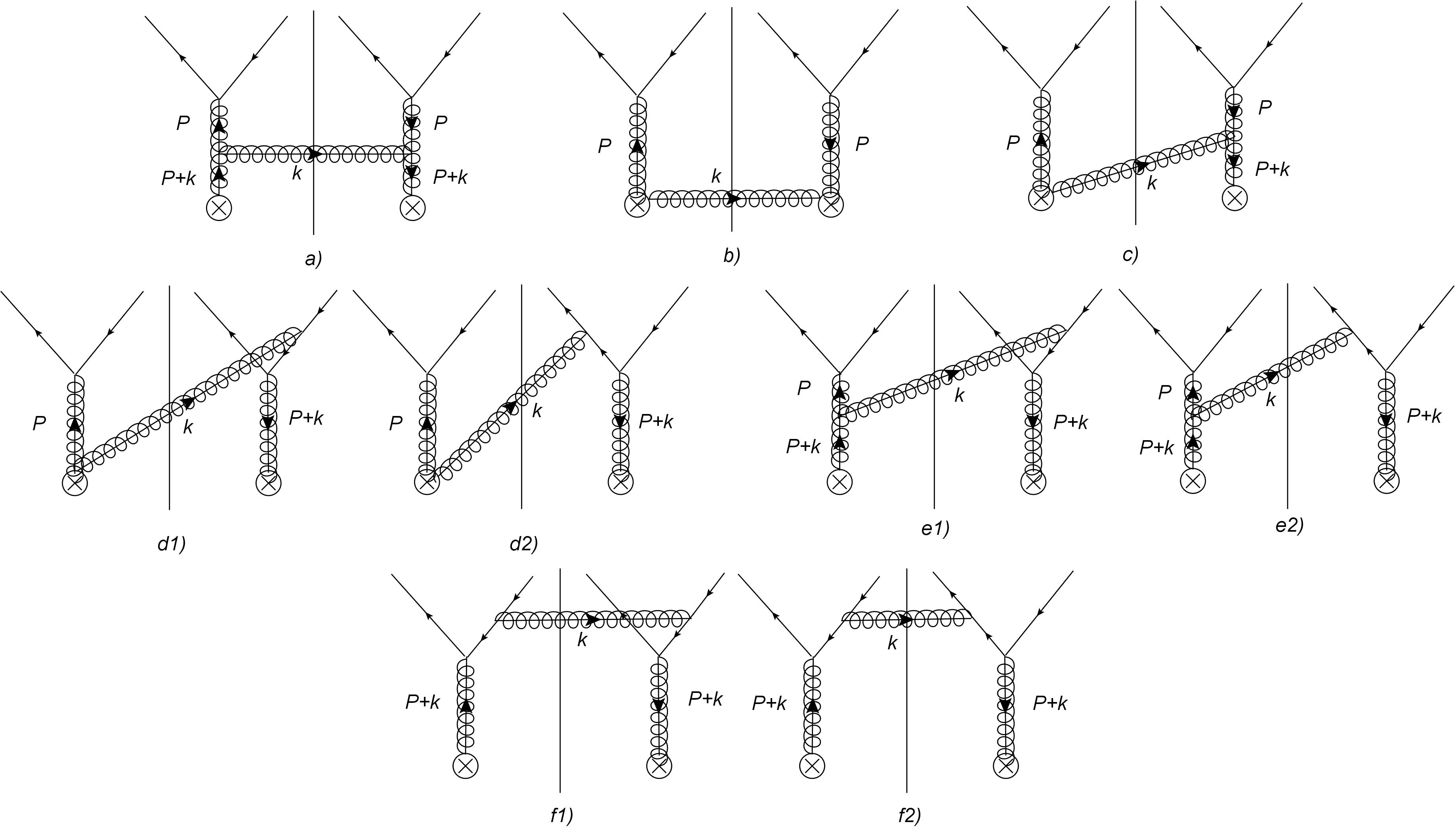}
    \caption{ Real diagrams contributing to NLO. Only diagrams c, d1 and d2 have rapidity divergences. Hermitian conjugates of diagrams c, d1, d2, e1, e2, f1 y f2 are not shown.}
    \label{fig:NLOreal}
\end{figure}

\section{Appendix} \label{sec:Calculations}

In this section we present the details for the calculation of the short-distance matching coefficient of the unpolarized gluon TMDFF at NLO into heavy quarkonium in the color-octet $^3S_1$ channel, which is shown in \eqref{eq:NLOtotal}. 
We have used dimensional regularization for ultraviolet (UV) divergences in the \textoverline{MS}-scheme, i.e. $\mu^2 \rightarrow \mu^2 e^{\gamma_E}/4 \pi$, and $\delta$-regularization for infrared (IR) and rapidity divergences. Our goals are to obtain the short-distance coefficient $d^{\rm{NLO}}_{g \rightarrow J/\psi}$ around the threshold by matching both sides of \eqref{eq:TMDFFfactorization}, to show the cancellation of the spurious rapidity divergences and to extract the Wilson matching coefficients of the TMDFF onto its integrated function, as shown in \eqref{eq:OPE}.

With $\delta$-regulator the relevant Feynman rules for the collinear gluon field in \eqref{eq:Bfield}, needed in the calculation at NLO, become
\begin{equation}
\begin{aligned}
\delta^{ab} \mathcal{B}_{n \perp}^{(0) \mu \nu} (k)  & =  \delta^{ab} \left(g_\perp^{\mu \nu} - \frac{k_\perp^\mu \bar{n}^\nu}{k^+ - i \delta^+} \right) ,\\
ig_s f^{cab} \mathcal{B}_{n \perp}^{(1) \mu \nu_1 \nu_2} (k_1,k_2) & = i g_s f^{cab} \left[ \frac{g_\perp^{\mu \nu_2} \bar{n}^{\nu_1}}{k_1^+-i \delta^+} - \frac{g_\perp^{\mu \nu_1} \bar{n}^{\nu_2}}{k_2^+-i \delta^+} \right. \\
& \left. 
+ \left( \frac{k_{1 \perp}^\mu }{k_2^+ - i \delta^+} - \frac{k_{2 \perp}^\mu}{k_2^+ - i \delta^+} \right) \frac{\bar{n}^{\nu_1} \bar{n}^{\nu_2}}{[(k_1^+ + k_2^+) - i 2 \delta^+]} \right] 
\,.
\end{aligned}
\end{equation}
where the superscripts $(0)$ and $(1)$ denote the power of the strong coupling constant $g$.
The next order is zero.

We assume that the relative momentum of the heavy-quark pair is small compared with the mass $m_c$, so   
$$P^2 = 4 \, E_q^2 = 4  (m_c^2 + \mathbf{q}^2) \simeq 4 \, m_c^2 = M^2\,.$$ 
The Dirac spinors for the $c$ an $\bar c$ in the frame in which the pair has total momentum $P$ are
\begin{equation}
\begin{aligned}
u (p_1) & = \frac{2 E_q + \Slash P \gamma_0}{\sqrt{4 E_q(P_0 + 2 E_q)(E_q + m_c )}} \left( (E_q+m_c) \, \xi, \, \mathbf{q} \cdot \mathbf{\sigma} \, \xi \right)^T, \\
v(p_2) & = \frac{2 E_q + \Slash P \gamma_0}{\sqrt{4 E_q(P_0 + 2 E_q)(E_q + m_c )}} \left( - \mathbf{q} \cdot \mathbf{\sigma} \, \xi, \, (E_q+m_c) \, \xi \right)^T,
\end{aligned}
\end{equation}
where $\xi$ and $\eta$ are the Pauli spinors and they are normalized such that $\xi^\dagger \xi = 1 $ and $\eta^\dagger \eta = 1$.\\

First, we calculate the left hand side of \eqref{eq:TMDFFfactorization}, that is the matrix element defined in \eqref{eq:TMDFFnorenor}, around the threshold ($\mathbf{q} = 0 $). To do this, we need the following nonrelativistic expansions in powers of $\mathbf{q}$, which we will obtain when we make the calculation of equation \eqref{eq:TMDFFfactorization} and which will allow us to compare with the LDMEs when matching:
\begin{equation}
\begin{aligned} \label{eq:Thres.Expansion}
\bar u (p_1) v(p_2) |_{\mathbf{q} = 0}  & =   0 ,\\
\bar u(p_1) \gamma^\mu v(p_2) |_{\mathbf{q} = 0}  & =  2 m_c \, L^\mu_i \, \xi^\dagger \sigma^i \eta ,\\
\bar u (p_1) (\gamma^\mu \gamma^\nu - \gamma^\nu \gamma^\mu) v(p_2) |_{\mathbf{q} = 0} & = 2 \left( P^\mu L^\nu_i - P^\nu L^\mu_i \right) \xi^\dagger \sigma^i \eta, \\
\bar u (p_1) (\gamma^\mu \gamma^\nu \gamma^\lambda - \gamma^\lambda \gamma^\nu \gamma^\mu) v(p_2) |_{\mathbf{q} = 0} & = - m_c \, L^\mu_i L^\nu_j L^\lambda_k \, \xi^\dagger \{ [\sigma^i, \sigma^j], \sigma^k\} \eta ,
\end{aligned}
\end{equation}
where $\sigma$ denotes the Pauli matrices and  $L^\mu_i$ is a boost matrix defined as
\begin{equation}
\begin{aligned} \label{eq:Lrelations}
g_{\mu \nu} L^\mu_i L^\nu_j & = - \delta^{ij},\\
L^\mu_i L^\nu_i & = - g^{\mu \nu} + \frac{P^\mu P^\nu}{P^2} .
\end{aligned}
\end{equation}

For example, equation \eqref{eq:TMDFFnorenor} at Leading Order (LO), which is shown in fig.~\ref{fig:NLOvirtual}, is
\begin{equation}
\begin{aligned}
\Delta_{g \rightarrow J/\psi}^{\rm{LO}} & = \frac{4 \pi \alpha_s \delta(1-z)}{8 (d-2)} \\
& \times \left[ \bar{u} (p_1) ( \gamma^{\tau_1} T^{a}_{ij}) v(p_2) \right] \left[ \bar{v} (p_2) ( \gamma^{\tau_2} T^{a}_{kl}) u(p_1) \right]\mathcal{B}_{n \perp}^{(0) \lambda \tau_1} (-P) \mathcal{B}_{n \perp}^{(0) \sigma \tau_2}(P) \frac{(-g_\perp^{\tau_1 \tau_2})}{M^4} 
\end{aligned}
\end{equation}
By using the expressions of \eqref{eq:Thres.Expansion} we get
\begin{equation}
\begin{aligned}
\left[\bar u(p_1) \gamma^{\tau_1} v(p_2) \right]  \left[\bar v(p_2) \gamma^{\tau_2} u(p_1) \right] & = 4m_c^2 \, L^{\tau_1}_i L^{\tau_2}_j \; \xi^\dagger \sigma^i \eta \times \eta^\dagger \sigma^j \xi .
\end{aligned}
\end{equation}
We can average that factor over rotations if we also average the projection operator on the right side of \eqref{eq:TMDFFfactorization}. 
The average of the spin factor is
\begin{gather}
    \overline{\xi^\dagger \sigma^i \eta \times \eta^\dagger \sigma^i \xi} = \frac{\delta^{ij}}{d-1} \xi^\dagger \sigma^k \eta \times \eta^\dagger \sigma^k \xi,
\end{gather}
so, by using the second relation in \eqref{eq:Lrelations} we get
\begin{equation}
\begin{aligned} \label{eq:GG}
\left[\bar u(p_1) \gamma^{\tau_1} v(p_2) \right]  \left[\bar v(p_2) \gamma^{\tau_2} u(p_1) \right] & = \frac{M^2}{d-1} \mathcal{L}^{\tau_1 \tau_2} \, \xi^\dagger \sigma^k \eta \times \eta^\dagger \sigma^k \xi ,
\end{aligned}
\end{equation} 
where we have defined
\begin{equation}
\begin{aligned} \label{eq:LL}
\mathcal{L}^{\mu \nu} \equiv - g^{\mu \nu} + \frac{P^\mu P^\nu}{M^2}.
\end{aligned}
\end{equation}
Therefore, the left hand side of \eqref{eq:TMDFFfactorization} at LO is
\begin{equation}
\begin{aligned} \label{eq:DLO}
D^{\rm{LO}}_{g \rightarrow J/\psi} (z) & = \frac{\pi \alpha_s}{8(d-1) m_c^2}  \delta(1-z) \, \xi^\dagger T^a  \sigma^k \eta \times \eta^\dagger T^a \sigma^k \xi
\end{aligned}
\end{equation}
After calculating that side of \eqref{eq:TMDFFfactorization} we note that we obtain the spin factor $\xi^\dagger T^a \sigma^k \eta  \times \eta^\dagger T^a \sigma^k \xi$ that defines the color and angular momentum configuration $n = \tensor[^3]{S}{_1^{[8]}}$. 
We now focus on the NRQCD side of the matching equation. 
The spinor structure $\xi^\dagger T^a \sigma^k \eta  \times \eta^\dagger T^a \sigma^k \xi$ can be defined as the expansion to LO in $\alpha_s$ of the following NRQCD matrix element:
\begin{equation}
\begin{aligned} \label{eq:OpNRQCD1}
\sandwich{0}{\chi^\dagger \sigma^k T^a \psi \mathcal{P}_{J/\psi} \psi^\dagger \sigma^k T^a \chi }{0}|_{\rm{pNRQCD}} & \simeq 4 m_c^2 \, \xi^\dagger T^a \sigma^k \eta \times \eta^\dagger T^a\sigma^k \xi ,
\end{aligned}
\end{equation}
and, in turn,
\begin{equation}
\begin{aligned} \label{eq:OpNRQCD2}
\braket{\mathcal{O}^8 (^3 S_1)} & = \frac{1}{4 m_c} \sandwich{0}{\chi^\dagger \sigma^k T^a \psi \mathcal{P}_H \psi^\dagger \sigma^k T^a \chi }{0} .
\end{aligned}
\end{equation}

Finally, by matching between the two sides of equation \eqref{eq:TMDFFfactorization} we can obtain the short-distance matching coefficient. 
We can conclude that by following the equations \eqref{eq:OpNRQCD1} and \eqref{eq:OpNRQCD2}, the short-distance matching coefficient will be the result of the left-hand side without the spin factor $\xi^\dagger T^a \sigma^k \eta  \times \eta^\dagger T^a \sigma^k \xi$ and with a $1/m_c$ factor:
\begin{equation}
\begin{aligned}
d^{\rm{LO}}_{g \rightarrow J/\psi} (z) =  \frac{\pi \alpha_s}{8(d-1) m_c^3}  \delta(1-z) 
\,.
\end{aligned}
\end{equation}

\subsection{Details for the calculation of the virtual contribution}
Diagrams contributing to the gluon TMDFF at NLO are shown in figures~\ref{fig:NLOvirtual} and~\ref{fig:NLOreal}. 
In this section we calculate the diagrams shown in fig.~\ref{fig:NLOvirtual} which are the so-called virtual diagrams. 
The diagram \ref{fig:NLOvirtual}a is the propagator correction for a gluon with invariant mass $M^2 =  4\, m_c^2$, diagrams \ref{fig:NLOvirtual}b1 and b2 describe the vertex correction factor and diagrams \ref{fig:NLOvirtual}c1 and c2 are the wave renormalization factor (WFR) for a heavy quark and anti-quark:

\begin{equation}
\begin{aligned}
    \Delta_{g \rightarrow J/\psi}^{\ref{fig:NLOvirtual} a,q} & = \frac{i g^4 M^2}{8 (d-2) (d-1)} \frac{\delta(1-z)}{2} \int_k \left[ \bar u(p_1) T^a_{ij} \gamma^{\tau_1} v(p_2) \right] \left[ \bar v(p_2)  T^a_{kl} \gamma^{\tau_2} u(p_1) \right] \\
    & \times  \frac{B_{n \perp}^{(0) \lambda \rho} (-P) B_{n \perp}^{(0) \sigma \tau_2} (P) (-g_\perp^{\lambda \sigma})  }{(P^2)^3 [k^2 + i 0] [(P-k)^2 + i 0]} \,  \text{Tr} \left[ \gamma^\rho \Slash k \gamma^{\tau_1} (\Slash P - \Slash k) \right]\\
    & = \frac{-i  g^4  \delta(1-z)  }{8 (d-2) (d-1) M^4}\int_k \frac{  p^+ ((d-2)k^- p^+ + 2 (d-4) \mathbf{k}_\perp^2) + (d-2) k^+ (M^2 - 2 k^- p^+) }{ [\mathbf{k}_\perp^2 - k^- k^+] [(k^+ - p^+)(M^2 - k^- p^+)+\mathbf{k}_\perp^2 p^+]} \\
    & = \frac{\pi \alpha_s \delta(1-z)}{8(d-1)m_c^2} \frac{\alpha_s}{\pi} \left[ - \frac{1}{6 } \frac{1}{\veuv} - \frac{1}{6} \ln \frac{\mu^2}{M^2} - \frac{5}{18} + \frac{i \pi}{6} \right] \xi^\dagger T^a \sigma^k \eta \times \eta^\dagger T^a \sigma^k \xi ,
\end{aligned}
\end{equation}

\begin{equation}
\begin{aligned}
    \Delta_{g \rightarrow J/\psi}^{\ref{fig:NLOvirtual}a,g} & = \frac{i g^4 M^2 C_A/2}{8 (d-2) (d-1)} \delta(1-z) \int_k \left[ \bar u(p_1) T^a_{ij} \gamma^{\tau_1} v(p_2) \right] \left[ \bar v(p_2)  T^a_{kl} \gamma^{\tau_2} u(p_1) \right] \\
    & \times  \frac{B_{n \perp}^{(0) \lambda \rho} (-P) B_{n \perp}^{(0) \sigma \tau_2} (P) (-g_\perp^{\lambda \sigma})  }{(P^2)^3 [k^2 + i 0] [(P-k)^2 + i 0]} \,  V^{\alpha \delta \rho} (-k,k-P,P) V^{\alpha \delta \tau_1}(k,-k+P,-P) \\
    & = \frac{-i  g^4 C_A/2 \,  \delta(1-z)  }{8 (d-2) (d-1) M^4}\int_k \frac{  1 }{ [\mathbf{k}_\perp^2 - k^- k^+] [(k^+ - p^+)(M^2 - k^- p^+)+\mathbf{k}_\perp^2 p^+]} \\
    & \times \left( p^+ ((d-2)k^- p^+ + 2 (3d-5) \mathbf{k}_\perp^2 - 5(d-2)M^2) + (d-2) k^+ (M^2 - 2 k^- p^+) \right) \\
    & = \frac{\pi \alpha_s \delta(1-z)}{8(d-1)m_c^2} \frac{\alpha_s C_A}{\pi} \left[  \frac{19}{48 } \frac{1}{\veuv}  + \frac{19}{48} \ln \frac{\mu^2}{M^2} + \frac{29}{36} - \frac{i \pi 19}{48} \right] \xi^\dagger T^a \sigma^k \eta \times \eta^\dagger T^a \sigma^k \xi ,
\end{aligned}
\end{equation}

\begin{equation}
\begin{aligned}
    \Delta_{g \rightarrow J/\psi}^{\ref{fig:NLOvirtual}a,G} & = \frac{i g^4 M^2 C_A}{8 (d-2) (d-1)} \delta(1-z) \int_k \left[ \bar u(p_1) T^a_{ij} \gamma^{\tau_1} v(p_2) \right] \left[ \bar v(p_2)  T^a_{kl} \gamma^{\tau_2} u(p_1) \right] \\
    & \times  \frac{B_{n \perp}^{(0) \lambda \rho} (-P) B_{n \perp}^{(0) \sigma \tau_2} (P) (-g_\perp^{\lambda \sigma})  }{(P^2)^3 [k^2 + i 0] [(P-k)^2 + i 0]} \, (p-k)^\rho k^{\tau_1} \\
    & = \frac{i g^4 C_A \delta(1-z)}{8 (d-2) (d-1) M^4} \int_k \frac{\mathbf{k}_\perp^2 \, p^+}{[\mathbf{k}_\perp^2 - k^+ k^-] [(k^+ - p^+) (M^2 -  k^- p^+) + \mathbf{k}_\perp^2 p^+]} \\
    & = \frac{\pi \alpha_s \delta(1-z)}{8(d-1) m_c^2} \frac{\alpha_s C_A}{\pi} \left[ \frac{1}{48} \frac{1}{\veuv} + \frac{1}{48} \ln \frac{\mu^2}{M^2} - \frac{1}{72} + \frac{i \pi}{48} \right] \xi^\dagger T^a \sigma^k \eta \times \eta^\dagger T^a \sigma^k \xi ,
\end{aligned}
\end{equation}

\begin{equation}
\begin{aligned}
    \Delta^{\ref{fig:NLOvirtual} a + h.c.}_{g \rightarrow J/\psi} & = 2\, \text{Re} \left[ n_f \Delta^{\ref{fig:NLOvirtual} a, q}_{g \rightarrow J/\psi} + \Delta^{\ref{fig:NLOvirtual} a, g}_{g \rightarrow J/\psi} + \Delta^{\ref{fig:NLOvirtual} a, G}_{g \rightarrow J/\psi} \right] \\
    & = \frac{\pi \alpha_s \delta(1-z)}{8(d-1) m_c^2} \, \frac{\alpha_s C_A}{\pi} \left[ \left( \frac{5}{6} - \frac{n_f}{3 C_A} \right) \left( \frac{1}{\veuv} + \ln \frac{\mu^2}{M^2}  \right) + \frac{19}{12} - \frac{5 n_f }{9 C_A}  \right] \\
    & \times \xi^\dagger T^a \sigma^k \eta \times \eta^\dagger T^a \sigma^k \xi \\
    & = \frac{\pi \alpha_s \delta(1-z)}{8(d-1) m_c^2} \, \frac{\alpha_s C_A}{\pi} \left[ \left( \frac{\beta_0}{2 C_A} -1  \right) \left( \frac{1}{\veuv} + \ln \frac{\mu^2}{M^2}  \right) + \frac{19}{12} - \frac{5 n_f }{9 C_A}  \right] \\
    & \times \xi^\dagger T^a \sigma^k \eta \times \eta^\dagger T^a \sigma^k \xi ,
\end{aligned}
\end{equation}

\begin{equation}
\begin{aligned}
    \Gamma^{a,\tau_1 (\ref{fig:NLOvirtual} b1)}_{ij } & = g^3 \mu^{2\ve} T^a_{ij} \frac{C_A}{2} \int_k \gamma^{\rho_1} \frac{\Slash P /2 - \Slash k + m_c}{k^2 - k \cdot P} \gamma^{\rho_2} \frac{V^{\tau_1\phantom{\rho_1} \phantom{\rho_2}}_{\phantom{\tau_1}\rho_1 \rho_2} (P,-k,-P+k)}{k^2 (P-k)^2} \\
    & = g^3 \mu^{2\ve} T^a_{ij} \frac{C_A}{2} \int_0^1 dx_2 \int_0^{1- x_2} dx_3 \frac{i}{2 (4\pi)^{2-\ve}} \left[ 4 (2 \ve -3) \frac{\gamma^{\tau_1} \Gamma(\ve)}{m^{2 \ve} ((2x_2+x_3)^2 - 4 x_2)^\ve} \right. \\
    &  + \left. \left(  2m(2 \Slash P \gamma^{\tau_1} - \gamma^{ \tau_1} \Slash P) + 4 m^2 (2x_2 +x_3 -1)^2 \gamma^{\tau_1}  \right. \right. \\
    & \left. \left. + 2 (1- \ve)p^{\tau_1} \Slash P (2x_2 + x_3 -1)^2  \right) \frac{\Gamma(1+\ve)}{m^{2+2\ve} ((2x_2+x_3)^2 - 4x_2)^{1+\ve}} \right] \\
    & = - i g C_A T^{a}_{ij} \frac{ \alpha_s}{\pi} \left[\gamma^{\tau_1} \left( \frac{3}{8} \frac{1}{\veuv}  + \frac{3}{8} \ln \frac{\mu^2}{M^2} + \frac{1}{3} \ln \, 2 + \frac{2}{3} + \frac{i 5\pi}{24}  \right) \right. \\
    & \left. + \frac{3}{m_c} \, \Slash P \gamma^{\tau_1}  \left( \frac{1}{8} \ln \, 2 - \frac{i \pi}{16} \right) \right] ,
\end{aligned}
\end{equation}

\begin{equation}
\begin{aligned}
\Gamma^{a,\tau_1 (\ref{fig:NLOvirtual} b2)}_{ij} & = - g^3 \mu^{2 \ve} T^a_{ij} \left( C_F - \frac{C_A}{2} \right)  \int_k \, \frac{1}{k^2} \, \gamma^\rho \frac{\Slash k + \Slash P /2 + m_c}{ k^2  + k \cdot P } \gamma^{\tau_1} \frac{\Slash k - \Slash P /2 + m_c}{ k^2  - k \cdot P } \gamma_\rho \\
& = - g^3 \mu^{2 \ve} T^a_{ij} \left( C_F - \frac{C_A}{2} \right) \int_0^1 dx_2 \int_0^{1- x_2} dx_3 \frac{i}{(4 \pi)^{2-\ve}} \left[ \gamma^{\tau_1} \frac{2(1-\ve)^2 \Gamma(\ve)}{(m_c (x_2-x_3))^{2\ve}} \right. \\
& \left. + \left( -2(1-\ve)m_c^2 ((x_2-x_3)^2 - 2) \gamma^{\tau_1} + 2 \ve m_c \Slash P \gamma^{\tau_1} \right) \frac{\Gamma(1+ \ve)}{(m_c (x_2-x_3))^{2+2\ve}} \right] \\
& = - i g \left( C_F - \frac{C_A}{2} \right) T^a_{ij} \frac{\alpha_s}{\pi} \left[ \gamma^{\tau_1} \left( \frac{1}{4} \frac{1}{\veuv} + \frac{1}{2} \frac{1}{\veir} + \frac{3}{4} \ln \frac{\mu^2}{M^2} \right. \right. \\ 
&\left. \left. + \frac{3}{2} \ln \, 2 - \frac{11}{4} - \frac{i \pi}{4} \right) + \frac{1}{4m_c} \Slash P \gamma^{\tau_1} \right],
\end{aligned}
\end{equation}

\begin{equation}
\begin{aligned}
   - i \Sigma^{(\ref{fig:NLOvirtual} c1)} & = - g^2 \mu^{2 \ve} C_F \int_k \frac{1}{k^2} \gamma^\mu \frac{\Slash p_1 + \Slash k + m_0}{(p_1 + k) - m_0^2} \gamma_{\mu} \\
   & = - g^2 \mu^{2\ve} C_F \int_k \frac{(2-d)(\Slash p_1 + \Slash k)^2 + d \, m_0}{k^2 [(p_1 + k)^2 - m_0^2 ]} \\
   & = - g^2 \mu^{2\ve} C_F \int_0^1 dx \int_k \frac{(2-d)(1-x) \Slash p_1 + d \, m_0}{[k^2 + p_1^2 x(1-x) - m_0^2 x]^2} ,
\end{aligned}
\end{equation}

In order to calculate the contribution of the renormalization of the heavy-quark wave function we need to define the following:
\begin{equation}
\begin{aligned}
    \Sigma(p_1^2) & = A(p_1^2) m_0 + B(p_1^2) \Slash p_1 , \\
    A(p_1^2) &  = \frac{4 \pi \alpha_s C_F}{(4 \pi)^{2-\epsilon}} \mu^{2 \ve} \Gamma(\ve) (4-2 \ve) \int_0^1 dx \left( m_0^2 x - p_1^2 x(1-x) \right)^{-\ve} , \\
    B(p_1^2) & = - \frac{4 \pi \alpha_s C_F}{(4 \pi)^{2-\ve}} \mu^{2 \ve} \Gamma(\ve) 2(1 -\ve) \int_0^1 dx (1-x) \left( m_0^2 x - p_1^2 x(1-x) \right)^{-\ve} 
\,.
\end{aligned}
\end{equation}
The renormalization factor of the heavy-quark wave function is then
\begin{equation}
\begin{aligned}
Z_Q - 1 & = B(m_c^2) + 2 m_c^2 \left. \frac{d(A+B)}{dp_1^2} \right|_{p_1^2 = m_0^2 = m_c^2} \\
& = - C_F \frac{\alpha_s}{\pi} \left[ \frac{1}{4} \frac{1}{\veuv} + \frac{1}{2} \frac{1}{\veir} + \frac{3}{4} \ln \frac{\mu^2}{M^2} + \frac{3}{2} \ln \, 2 + 1 \right]
\,.
\end{aligned}
\end{equation}
The UV pole comes from $B(m_c^2)$ and the IR pole from the other term. \\

Diagrams \ref{fig:NLOvirtual}(d1+d2) and its hermitian conjugate give
\begin{equation}
\begin{aligned}
\Delta^{\ref{fig:NLOvirtual}(d1+d2)+h.c.}_{g \rightarrow J/\psi} = & \frac{i g^4 C_A}{32(1-\ve)} \delta(1-z) \int_k \left[ - \bar u (p_1) \gamma^{\rho} T^a_{ij} \left( \frac{(-\Slash p_1 + \Slash k + m_c)}{(p_1-k)^2 - m_c^2} \right) \gamma^{\tau_1} v(p_2)  \right. \\
+ & \left. \bar u (p_1)  \gamma^{\tau_1}   T^a_{ij} \left( \frac{(-\Slash p_2 + \Slash k + m_c)}{(p_2-k)^2 - m_c^2} \right) \gamma^{\rho}   v (p_2)\right] \left[ \bar v (p_2) (\gamma^{\tau_2} T_{kl}^{a} ) v(p_1) \right]\\
\times & \frac{1}{P^2[k^2+i0][(P-k)^2+i0]} B_{n \perp \lambda \tau_1 \rho}^{(1)}(-P,-k) B_{n \perp}^{(0) \sigma \tau_2}(P) (-g_\perp^{\lambda \sigma}) + h.c. \\
= & - \frac{\alpha_s^2 C_A}{8(d-1)m_c^2} \delta(1-z) \, 32 \pi^2 m_c^2 \, \text{Im}  \left( P^+ I_{ABCD} - I_{ABD} \right) \\
&\times 
\xi^\dagger T^a \sigma^k \eta \times \eta^\dagger T^a \sigma^k \xi\\
= &  \frac{\alpha_s^2 C_A}{8(d-1) m_c^2} \delta(1-z) \left[  - \frac{1}{2} \ln^2 \delta + 2 \ln \, 2 - \frac{13 \pi^2}{24} \right] \\
&\times 
\xi^\dagger T^a \sigma^k \eta \times \eta^\dagger T^a \sigma^k \xi .
\end{aligned}
\end{equation}
with
\begin{equation}
\begin{aligned}
\int_k & \equiv \int \frac{d^{4-2 \ve} k}{(2 \pi)^{4 - 2 \ve}} .
\end{aligned}
\end{equation}
Here, we have a different spinorial structures from the previous one:
\begin{equation}
\begin{aligned}
\left[ \bar u (p_1) \gamma^{\tau_1} \gamma^{\rho}  v(p_2) \right] \left[ \bar v (p_2)    \gamma^{\tau_2}   u (p_1) \right] = & \frac{2 m_c}{d-1} \left( P^{\tau_1} \mathcal{L}^{\rho \tau_2} - P^{\rho} \mathcal{L}^{\tau_1 \tau_2} \right) \xi^\dagger \sigma^k \eta \times \eta^\dagger \sigma^k \xi, \\
\left[ \bar u (p_1) \gamma^{\tau_1} \gamma^{\mu} \gamma^\rho  v(p_2) \right] \left[ \bar v (p_2)   \gamma^{\tau_2}   u (p_1) \right] = & \frac{M^2}{d-1} \left( g^{\mu \rho} \mathcal{L}^{\tau_1 \tau_2} - g^{\tau_1 \rho} \mathcal{L}^{\mu \tau_2} + g^{\tau_1 \mu} \mathcal{L}^{\rho \tau_2} \right) \\
& \xi^\dagger \sigma^k \eta \times \eta^\dagger \sigma^k \xi,
\end{aligned}
\end{equation}
where we have used that the rotational average of $\xi^\dagger \{ [\sigma^i, \sigma^k], \sigma^l\} \eta \times \eta^\dagger \sigma^j  \xi$ vanishes.\\
Diagrams \ref{fig:NLOvirtual}e and its hermitian conjugate give
\begin{equation}
\begin{aligned}
\Delta^{\ref{fig:NLOvirtual}e+h.c.}_{g \rightarrow J/\psi}  & = \frac{g^4 C_A }{16(1-\ve)} \int_k   \left[ \bar{u} (p_1) ( \gamma^{\tau_1} T^{a}_{ij}) v(p_2) \right] \left[ \bar{v} (p_2) ( \gamma^{\tau_2} T^{a}_{kl}) u(p_1) \right] \\
& \times \frac{B_{n \perp \lambda \beta \rho}^{(1)}(-P,-k) B_{n \perp}^{(0) \sigma \tau_2}(P) V^{\rho \beta \tau_1}(k,P-k,-P)(-g_\perp^{\lambda \sigma})}{P^2 [k^2+i0][(P-k)^2+i0]} + h.c. \\
& = -  \frac{\alpha_s^2 C_A}{8(d-1) m_c^2} \delta(1-z) \, 8 \pi^2 \, \text{Im} \left( 2 P^+ I_{ABC} - I_{AB} \right) \xi^\dagger T^a \sigma^k \eta \times \eta^\dagger T^a \sigma^k \xi \\
& = \frac{\alpha_s^2 C_A}{8(d-1) m_c^2} \delta(1-z) \left[ \frac{1}{2 \veuv} \left(1 + 2 \ln \, \delta \right) - \frac{1}{2} \ln^2 \delta + \ln \, \delta \; \ln \frac{\mu^2}{M^2} \right. \\
& \left. + \frac{1}{2} \ln \frac{\mu^2}{M^2}  +1  - \frac{5 \pi^2}{24} \right] \xi^\dagger T^a \sigma^k \eta \times \eta^\dagger T^a \sigma^k \xi,
\end{aligned}
\end{equation}
where $V^{\alpha \beta \gamma}(p,q,k) = g^{\alpha \gamma}(k-p)^\beta + g^{\beta \gamma}(q-k)^\alpha + g^{\alpha \beta} (p-q)^\gamma$ and we have used eq.~\eqref{eq:GG}.

The integrals that appear in the calculation of the virtual contribution are as follows
\begin{equation}
\begin{aligned}
I_{ABC} & = \int \frac{d^{4-2\ve} k}{(2 \pi)^{4-2 \ve}} \frac{1}{[k^2 + i 0] [(k-P)^2 + i 0] [(P-k) \cdot n + i \delta^+]} \\
& = \frac{-i}{16 \pi^2 P^+} \left[ \left( \frac{1}{\veuv} + \ln \frac{\mu^2}{M^2} + i \pi \right) \left(  \ln \, \delta + \frac{i \pi}{2} \right) \right. \\
& \left. - \frac{1}{2} \ln^2 \delta + \frac{7 \pi^2}{24} - \frac{i3 \pi}{2} \ln \, \delta  \right] + \mathcal{O}(\ve) , \\
I_{ACD} & = \int \frac{d^{4-2\ve} k}{(2 \pi)^{4 -2 \ve}} \frac{1}{[k^2+ i 0][(P-k)\cdot n + i \delta^+]}  \frac{1}{[(k-P/2)^2-m_c^2+i 0]} \\
& = \frac{-i}{16 \pi^2 P^+} \left[ - \frac{2 \ln \, 2}{\veuv} - 2 \ln \, 2 \, \ln \frac{\mu^2}{M^2} - 2 \ln^2 2 - \frac{\pi^2}{3} \right] + \mathcal{O}(\ve),\\
I_{BCD} & = \int \frac{d^{4 -2 \ve} k}{(2 \pi)^{4 -2 \ve}} \frac{1}{[(k-P)^2+i 0][(P-k) \cdot n + i \delta^+]}  \frac{1}{[(k-P/2)^2 + i 0]}, \\
& = \frac{-i}{16 \pi^2 P^+} \left[ \left( \frac{1}{\veuv} + \ln \frac{\mu^2}{M^2} + i \pi \right) \left( 2 \ln \, \delta +2 \ln \, 2 + i \pi \right) \right. \\
& \left. - 2 \ln^2 \delta - \frac{\pi^2}{6} + 2 \ln^2 2 - i2  \pi \ln\, \delta \right] + \mathcal{O}(\ve) ,\\
I_{AB} & = \int \frac{d^{4-2 \ve} k}{(2 \pi)^{4-2\ve}} \frac{1}{[k^2+i0] [(k-P)^2+i0]} = \frac{i}{16 \pi ^2} \left[ \frac{1}{\veuv} +\ln{\frac{\mu ^2}{M^2}} +2 - i \pi \right] + \mathcal{O}(\ve),\\
I_{AD} & =  \int \frac{d^{4-2 \ve} k}{(2 \pi)^{4-2\ve}} \frac{1}{[k^2+i0] [(k-P/2)^2-m_c^2]} = \frac{i}{16 \pi ^2} \left[ \frac{1}{\veuv} +\ln{\frac{\mu ^2}{M^2}} +2+ 2 \ln \, 2 \right] + \mathcal{O}(\ve).
\end{aligned}
\end{equation}
and $4m_c^2 I_{ABCD} = I_{ACD} + I_{BCD} - 2 I_{ABC}$ and $4 m_c^2 I_{ABD} = 2 (I_{AD}-I_{AB})$.
\\

We note that we obtain the same spin factor $\xi^\dagger T^a \sigma^k \eta \times \eta^\dagger T^a \sigma^k \xi$ from all the virtual diagrams, which in turn, is the same as the one we obtain in eq.(\ref{eq:DLO}) at LO. As we said in the calculation of the LO contribution, this spin factor defines the configuration $n = \tensor[^3]{S}{_1^{[8]}}$. In order to obtain the SDC of each diagram, matching translates into removing the spin factor and adding the factor $1/m_c$ to the previous results:
\begin{equation}
\begin{aligned}
d^{\ref{fig:NLOvirtual}(a+b+c)+ h.c.}_{g \rightarrow J/\psi} (z)  & =  \frac{\alpha_s^2 C_A}{8 (d-1) m_c^3} \delta(1-z) \left[ \frac{1}{\veuv} \left( \frac{4}{3} - \frac{n_f}{3 C_A } \right) - \frac{1}{2 \veir} \right.\\
& \left. + \frac{41}{18} - \frac{5 n_f}{9 C_A} + \frac{2}{3} \ln \, 2 + \ln\frac{\mu^2}{M^2} \left( \frac{5}{6} - \frac{n_f}{3 C_A} \right) \right],\\
d^{\ref{fig:NLOvirtual}d + h.c.}_{g \rightarrow J/\psi} (z; \delta) & =  \frac{\alpha_s^2 C_A}{8(d-1) m_c^3} \delta(1-z) 
\left[ - \frac{1}{2} \ln^2 \delta + 2 \ln \, 2 - \frac{13 \pi^2}{24} \right],\\
d^{\ref{fig:NLOvirtual} e + h.c.}_{g \rightarrow J/\psi} (z; \delta) & =  \frac{\alpha_s^2 C_A}{8(d-1) m_c^3} \delta(1-z)
\left[ \frac{1}{2 \veuv} \left(1 + 2 \ln \, \delta \right) - \frac{1}{2} \ln^2 \delta + \ln \, \delta \; \ln \frac{\mu^2}{M^2} \right. \\
& \left. + \frac{1}{2} \ln \frac{\mu^2}{M^2}  +1  - \frac{5 \pi^2}{24} \right].
\end{aligned}
\end{equation}
Here $\ref{fig:NLOvirtual}b$ denotes $\ref{fig:NLOvirtual} b1 + \ref{fig:NLOvirtual} b2$ and so on.

\subsection{Details for the calculation of the real contribution}
In this section we calculate the contributon of the diagrams shown in fig.\ref{fig:NLOreal} which are the so-called real diagrams.\\
Diagram \ref{fig:NLOreal}a gives
\begin{equation}
\begin{aligned} \Delta_{g \rightarrow J/\psi}^{\ref{fig:NLOreal} a}  = &   \frac{2\pi g^4 C_A P^+ }{16 (1-\ve) } \int_k \left[ \bar{u} (p_1) ( \gamma^{\tau_1} T^{a}_{ij}) v(p_2) \right] \left[ \bar{v} (p_2) ( \gamma^{\tau_2} T^{a}_{kl}) u(p_1) \right] \frac{\delta(k^+ + P^+ (1-1/z))}{(P^2)^2 \, [(P+k)^2 + i 0]^2}  \\
\times & \delta(k^2) \theta(k^+)   B_{n \perp \lambda \alpha}^{(0)}(-P-k) B_{n \perp}^{(0) \lambda \beta}(P+k) V^{\alpha \rho \tau_1}(P+k,-k,-P) V_{\rho \beta \tau_2}(k,-P-k,P) \\
= & \frac{2^{2\varepsilon-3} \pi^{2\varepsilon -1} \alpha_s^2 C_A}{M^4(d-1)(1-\varepsilon) } \int d^{2-2 \varepsilon} \mathbf{k}_\perp \frac{z^{-3} (1-z)^{-1}}{\left[\mathbf{k}_\perp^2 + \frac{M^2 (1-z)}{z^2} \right]^2} \left[ (\mathbf{k}_\perp^2)^2 z^4 \left(z^2-z-\varepsilon +1\right) \right. \\
&-\mathbf{k}_\perp^2 z^4 M^2 (z-1) z^2 \left(z^3 (4 \varepsilon -5)+z^2 (6-4 \varepsilon )+z (4 \varepsilon -5)-2 \varepsilon +2\right) \\
& \left. +M^4 (z-1)^2 \left(z^2+4 z-1\right) (\varepsilon -1) \right] \; \xi^\dagger T^a \sigma^k \eta  \times \eta^\dagger T^a \sigma^k \xi
\end{aligned}
\end{equation}
In the second equality, we have used that
\begin{equation}
\begin{aligned}
\left[\bar u(p_1) \gamma^{\tau_1} v(p_2) \right]  \left[\bar v(p_2) \gamma^{\tau_2} u(p_1) \right] & = \frac{M^2}{d-1} \mathcal{L}^{\tau_1 \tau_2} \, \xi^\dagger \sigma^k \eta \times \eta^\dagger \sigma^k \xi ,
\end{aligned}
\end{equation}
where $\mathcal{L}^{\mu \nu}$ is defined in eq.(\ref{eq:LL}).\\
Diagram \ref{fig:NLOreal}b gives
\begin{equation}
\begin{aligned}\Delta_{g \rightarrow J/\psi}^{\ref{fig:NLOreal} b}  = & \frac{2 \pi g^4 C_A P^+}{16 (1-\ve)}  \int_k \left[ \bar{u} (p_1) ( \gamma^{\tau_1} T^{a}_{ij}) v(p_2) \right] \left[ \bar{v} (p_2) ( \gamma^{\tau_2} T^{a}_{kl}) u(p_1) \right] \frac{\delta(k^+ + p^+ (1-1/z))}{(p^2)^2}\\
\times & \delta(k^2) \theta(k^+) B_{n \perp \lambda \tau_1 \rho}^{(0)}(-p,-k) B_{n \perp}^{(0) \sigma \rho \tau_2}(k,p) (-g_\perp^{\lambda \sigma})\\
= &  \frac{2^{2\varepsilon - 3} \pi^{2 \varepsilon -1} \alpha_s^2 C_A}{M^4 (d-1)} \int d^{d-2} \mathbf{k}_\perp z (1-z)^{-1} \; \xi^\dagger T^a \sigma^k \eta  \times \eta^\dagger T^a \sigma^k \xi .
\end{aligned}
\end{equation}
Diagrams \ref{fig:NLOreal}c and its Hermitian conjugate give
\begin{equation}
\begin{aligned}\Delta_{g \rightarrow J/\psi}^{\ref{fig:NLOreal}c + c^*} = &  \frac{i 2 \pi g^4 C_A P^+}{16 (1-\ve)}  \int_k \left[ \bar{u} (p_1) (\gamma^{\tau_1} T^{c_1}_{ij}) v(p_2) \right] \left[ \bar{v} (p_2) ( \gamma^{\tau_2} T^{c_2}_{kl}) u(p_1) \right] \frac{\delta(k^+ + p^+(1-1/z))}{(p^2)^2 [(p+k)^2 + i0]}\\
\times &  \delta(k^2) \theta(k^+) B_{n \perp \lambda \tau_1 \rho}^{(1)}(-p,-k)  B_{n \perp}^{(0) \sigma \beta}(p+k) V_{\rho \beta \tau_2}(k,-p-k,p) (-g_\perp^{\lambda \sigma}) + h.c.\\
= & - \frac{2^{2\varepsilon-3} \pi^{2 \varepsilon -1} \alpha_s^2 C_A}{M^4 (d-1)(1- \varepsilon)} \int d^{d-2} \mathbf{k}_\perp \frac{z^{-1} (1-z)}{\left[ \mathbf{k}_\perp^2 + \frac{M^2 (1-z)}{z^2} \right] [(1-z)^2 + \delta^2 z^2 ] } \\
 & \times \left[ \mathbf{k}_\perp^2 z^2 (z^2 -z-2 \varepsilon +2) + 2 M^2(2 z^2 -z+1) (\varepsilon -1 ) \right] \; \xi^\dagger T^a \sigma^k \eta  \times \eta^\dagger T^a\sigma^k \xi .
\end{aligned} 
\end{equation}
Diagrams \ref{fig:NLOreal}d and its Hermitian conjugate give
\begin{equation}
\begin{aligned}\Delta_{g \rightarrow J/\psi}^{\ref{fig:NLOreal}(d1+d2)+h.c.} & = \frac{i \pi g^4 C_A P^+}{16 (1-\ve)} \int_k \left[ \bar u (p_1) (\gamma^{\tau_1} T_{ij}^{a} ) v(p_2) \right] \left[ - \bar v(p_2) \gamma^{\tau_2} T^a_{kl} \left( \frac{\Slash p_1 + \Slash k + m_c}{(p_1 + k)^2 - m_c^2} \right)  \gamma^\rho u(p_1) \right. \\
& \left. + \bar v (p_2)  \gamma^{\rho}  T^a_{kl} \left( \frac{-\Slash p_2 - \Slash k + m_c}{(p_2+k)^2 - m_c^2} \right) \gamma^{\tau_2}   u (p_1) \right] \frac{\delta(k^+ + P^+(1-1/z)) \delta(k^2) \theta(k^+)}{P^2 [(P+k)^2 + i0]} \\
\times &  B_{n \perp \lambda \tau_1 \rho}^{(1)}(-P,-k)  B_{n \perp}^{(0) \sigma \tau_2}(P+k) (- g_\perp^{\lambda \sigma})\\
= & \frac{(2 \pi)^{2 \varepsilon -2} \alpha_s^2 C_A}{M^2 (d-1) (1- \varepsilon)} \int d^{d-2} \mathbf{k}_\perp \frac{z^{-2} (1-z)^2}{\left[ \mathbf{k}_\perp^2 + \frac{M^2 (1-z)}{z^2} \right] \left[ \mathbf{k}_\perp^2 + \frac{M^2 (1-z)^2}{z^2} \right] \left[ (1-z)^2 + \delta^2 z^2 \right]} \\
& \times \left[ \mathbf{k}_\perp^2 z^2 (z^2 -z - \varepsilon +1) + M^2 (z^2 +1) (\varepsilon -1 ) \right] \; \xi^\dagger T^a_{ij} \sigma^k \eta  \times \eta^\dagger T^a_{kl} \sigma^k \xi .
\end{aligned}
\end{equation}
Here, we have a different spinorial structures from the previous ones:
\begin{equation}
\begin{aligned}
\left[ \bar u (p_1) \gamma^{\tau_1}  v(p_2) \right] \left[ \bar v (p_2)  \gamma^{\rho}  \gamma^{\tau_2}   u (p_1) \right]& = - \frac{2 m_c}{d-1} \left( P^\rho \mathcal{L}^{\tau_1 \tau_2} - P^{\tau_2} \mathcal{L}^{\tau_1 \rho} \right) \xi^\dagger \sigma^k \eta \times \eta^\dagger \sigma^k \xi, \\
\left[ \bar u (p_1) \gamma^{\tau_1}  v(p_2) \right] \left[ \bar v (p_2)  \gamma^{\rho} \gamma^\mu \gamma^{\tau_2}   u (p_1) \right] &= \frac{M^2}{d-1} \left( g^{\mu \tau_2} \mathcal{L}^{\tau_1 \rho} - g^{\rho \tau_2} \mathcal{L}^{\tau_1 \mu} + g^{\rho \mu} \mathcal{L}^{\tau_2} \right) \xi^\dagger \sigma^k \eta \times \eta^\dagger \sigma^k \xi,
\end{aligned}
\end{equation}
where we have used that the rotational average of $\xi^\dagger \sigma^i \eta \times \eta^\dagger \{ [\sigma^j, \sigma^k], \sigma^l\} \xi$ vanishes.\\
Diagrams \ref{fig:NLOreal}e1 + \ref{fig:NLOreal}e2 and its Hermitian conjugate give
\begin{equation}
\begin{aligned} \Delta_{g \rightarrow J/\psi}^{\ref{fig:NLOreal}(e1+e2)+ \rm{h.c.} } & =  \frac{\pi g^4 C_A P^+}{16(1-\ve)} \int_k \left[ \bar u (p_1) ( \gamma^{\tau_1} T_{ij}^{a} ) v(p_2) \right] \left[  \bar v(p_2) \gamma^{\tau_2} T^a_{kl} \left( \frac{\Slash p_1 + \Slash k + m_c}{(p_1 + k)^2 - m_c^2} \right)  \gamma^\rho u(p_1) \right. \\
& \left. - \bar v (p_2)  \gamma^{\rho}  T^a_{kl} \left( \frac{-\Slash p_2 - \Slash k + m_c}{(p_2+k)^2 - m_c^2} \right) \gamma^{\tau_2}   u (p_1) \right] \frac{\delta(k^+ + P^+(1-1/z)) \delta(k^2) \theta(k^+)}{P^2 [(P+k)^2 + i 0]^2} \\
\times &  B_{n \perp \lambda \alpha}^{(0)}(-P-k)  V^{\alpha \rho \tau_1}(P+k,-k,-P) B_{n \perp}^{(0) \lambda \tau_2}(P+k) + h.c.\\
= & \frac{2^{2\varepsilon - 3} \pi^{2 \varepsilon -1} \alpha_s^2 C_A}{M^2(d-1) (1- \varepsilon) } \int d^{2-2\varepsilon} \mathbf{k}_\perp \frac{z^{-4}}{\left[ \mathbf{k}_\perp^2 + \frac{M^2 (1-z)}{z^2} \right]^2 \left[ \mathbf{k}_\perp^2 + \frac{M^2 (1-z)^2}{z^2} \right] } \\
\times & \left[ (\mathbf{k}_\perp^2)^2 z^4 (3 (z-1) z-2 \varepsilon +2) \right. \\
+ &\mathbf{k}_\perp^2 M^2 (z-1) z^2 ((z (4 (z-1) z+5)-2) (4-2 \varepsilon )+((2-5 z) z-7) z+4)\\
+ & 2 M^4 (z-1)^2 (5 z-1) (\varepsilon -1)  \left. \right] \; \xi^\dagger T^a \sigma^k \eta  \times \eta^\dagger T^a \sigma^k \xi
\end{aligned}
\end{equation}
Diagrams \ref{fig:NLOreal}f1 + \ref{fig:NLOreal}f2 and its Hermitian conjugate give
\begin{equation}
\begin{aligned}\Delta_{g \rightarrow J/\psi}^{\ref{fig:NLOreal}(f1+f2)+ h.c.} & = - \frac{2\pi g^4 P^+}{16(1-\ve)} \int_k \left[ \bar u (p_1)  \gamma^{\tau_1} T^{a}_{ij} \left( \frac{ - \Slash p_2 - \Slash k + m_c}{(p_2 + k)^2 - m_c^2} \right)  \gamma^{\rho} v (p_2) \right]\\
\times &  \left[  C_F \, \bar v(p_2) \gamma^{\rho} T^{a}_{kl} \left( \frac{- \Slash p_2 - \Slash k + m_c)}{(p_2 + k)^2 - m_c^2} \right)  \gamma^{\tau_2}   u (p_1) \right. \\
& \left. +  (C_F - C_A/2) \, \bar v(p_2) \gamma^{\tau_2} T^a_{kl} \left( \frac{\Slash p_1 + \Slash k + m_c}{(p_1+k)^2 -m_c^2} \right) \gamma^\rho u(p_1) \right]\\
\times & \frac{\delta(k^+ + P^+(1-1/z))\delta(k^2) \theta(k^+)}{[(P+k)^2 + i0]^2}  B_{n \perp \lambda \tau_1}^{(0)}(-P-k) B_{n \perp}^{(0) \sigma \tau_2}(P+k) (-g_\perp^{\lambda \sigma}) + h.c.\\
= & \frac{2^{2\varepsilon - 3 } \pi^{2 \varepsilon -1 } \alpha_s^2 C_A}{(d-1)(1-\varepsilon)} \int d^{2-2\varepsilon} \mathbf{k}_\perp \frac{z^{-5} (1-z)}{\left[ \mathbf{k}_\perp^2 + \frac{M^2 (1-z)}{z^2} \right]^2 \left[ \mathbf{k}_\perp^2 + \frac{M^2 (1-z)^2}{z^2} \right]^2 } \\
\times & \left[ (\mathbf{k}_\perp^2)^2 z^4 \left(-2 z^2+2 z+\varepsilon -1\right) \right.\\
+ & 2 \mathbf{k}_\perp^2 M^2 (z-1) z^2 \left(z^3 (2 \varepsilon -3)-2 z^2 (\varepsilon -2)+z (3 \varepsilon -4)-\varepsilon +1\right) \\
+ & M^4 (z-1)^2 \left(z^2-6 z+1\right) (\varepsilon -1) \left. \right] \xi^\dagger T^a \sigma^k \eta \times \eta^\dagger T^a \sigma^k \xi
\end{aligned}
\end{equation}
Where we have used that
\begin{equation}
\begin{aligned}
& \left[ \bar u (p_1) \gamma^{\tau_1} \left( - \Slash p /2 - \Slash k +m \right)  \gamma^{\rho}  v (p_2) \right] \left[ \bar v(p_2) \gamma^{\rho} \left(- \Slash p /2 - \Slash k + m \right) \gamma^{\tau_2}  u (p_1) \right]   \\ 
& =  \frac{M^2}{(d-1)} \left( (-p/2-k)^{\rho} L^{\tau_1}_i - g^{\tau_1 \rho} (-p/2-k)_{\alpha} L^{ \alpha}_i + (-p/2-k)^{\tau_1} L^{\rho} + \frac{p^{\tau_1} L^{\rho}_i - p^{\rho} L^{\tau_1}_i}{2} \right)\\
& \times \left( (-p/2-k)^{\tau_2} L^{\rho}_i - g^{\tau_2 \rho} (-p/2-k)_{\beta} L^{\beta}_i + (-p/2-k)^\rho L^{\tau_2}_i - \frac{p^\rho L^{\tau_2} - p^{\tau_2} L^{\rho}_i}{2} \right) \\
& \xi^\dagger T^a \sigma^k \eta \times \eta^\dagger T^a \sigma^k \xi .
\end{aligned}
\end{equation}

Finally, we note that we obtain the same structure $\xi^\dagger T^a \sigma^k \eta \times \eta^\dagger T^a \sigma^k \xi$ for all diagrams. 
As we mentioned in the calculation at LO, this spin factor defines the configuration $n = ^3S_1^{[8]}$. 
In order to obtain the SDCs from each diagram, matching translates into removing the spin factor and adding the factor $1/m_c$ to the previous results.

\subsubsection{Fourier transform}
We show the calculation of the Fourier transform of the results of the previous section. We separate the calculation into the diagrams that have rapidity divergences and those that do not.\\

The real contribution of diagrams $c$ and $d$ as a function of $\mathbf{k}_\perp$ is
\begin{equation}
\begin{aligned} \label{eq:cdFT}
d^{\rm{c,d}}_{g \rightarrow J/\psi} (z, \mathbf{k}_\perp; \delta) & = \frac{4^{\varepsilon-1} \pi^{2\varepsilon -1} \alpha_s^2 C_A}{(d-1)(1-\varepsilon) M^5} \frac{z^{-3} (1-z)}{\left[\mathbf{k}_\perp^2 + \frac{M^2 (1-z)}{z^2} \right] \left[\mathbf{k}_\perp^2 + \frac{M^2 (1-z)^2}{z^2} \right] [(1-z)^2 + \delta^2 z^2]} \\
& \times \left[ (\mathbf{k}_\perp^2)^2 z^4 \left(-z^2+z+2 \varepsilon -2\right)+\mathbf{k}_\perp^2 M^2 z^3 \left(z^3-z^2+z (3-4 \varepsilon )+1\right) \right. \\
& \left. + 2 M^4 \left(z^4-4 z^3+4 z^2-2 z+1\right) (1- \varepsilon) \right] \, .
\end{aligned}
\end{equation}
We need the following integrals in order to obtain that contribution as a function of $\mathbf{b}_\perp$,
\begin{equation}
\begin{aligned}
I_n (z,\mathbf{b}_\perp) & = \int d^{d-2} \mathbf{k}_\perp e^{i  \mathbf{k}_\perp \cdot \mathbf{b}_\perp} \left( \frac{(\mathbf{k}_\perp^2 )^n}{\left[\mathbf{k}_\perp^2 + \frac{M^2 (1-z)}{z^2} \right] \left[\mathbf{k}_\perp^2 + \frac{M^2 (1-z)^2}{z^2} \right]} \right),
\end{aligned}
\end{equation}
\begin{equation}
\begin{aligned}
I_0 & = \frac{2 \pi^{1-\varepsilon} B^{\varepsilon/2} z^{\varepsilon+1}}{M^{\varepsilon+2}} \left( - \frac{K_\varepsilon (a \sqrt{1-z})}{(1 -z )^{1+\varepsilon/2}} + \frac{K_\varepsilon (a (1-z))}{(1-z)^{1 + \varepsilon}}  \right), \\
I_1 & = \frac{2 \pi^{1-\varepsilon} B^{\varepsilon/2} z^{\varepsilon-1}}{M^{\varepsilon}} \left( \frac{K_\varepsilon (a \sqrt{1-z})}{(1-z)^{\varepsilon/2}} - \frac{K_\varepsilon (a (1-z))}{(1 -z )^{\varepsilon-1}} \right),\\
I_2 & =  \frac{2 \pi^{1-\varepsilon} B^{\varepsilon/2} z^{\varepsilon-3}}{M^{\varepsilon-2}} \left( - \frac{K_\varepsilon (a \sqrt{1-z})}{(1-z)^{\varepsilon/2-1}} + \frac{K_\varepsilon (a (1-z))}{(1 -z )^{\varepsilon-3}} \right),
\end{aligned}
\end{equation}
where $B= \mathbf{b}_\perp^2/4$ and $a = 2 \sqrt{B M^2/z^2}$. \\
Therefore, the Fourier transform of the equation~\eqref{eq:cdFT} is the following
\begin{equation}
\begin{aligned} \label{eq:cdBT}
d^{\rm{c,d}}_{g \rightarrow J/\psi} (z, \mathbf{b}_\perp; \delta) & = - \frac{2^{2 \varepsilon -1} \pi^\varepsilon B^{\varepsilon/2} \alpha_s^2 C_A  }{M^{3 + \varepsilon}  (d-1) (1-\varepsilon) } \frac{z^{\varepsilon} (1-z)^{1-\varepsilon}}{[(1-z)^2 + \delta^2 z^2] }  \\
& \times \left[ (1-z)^{\frac{\varepsilon }{2}+1} (z-2 \varepsilon +1) K_{\varepsilon }\left(a \sqrt{1-z}\right) \right. \\
& \left. - \frac{2 ((z-1) z ((z-2) z-2 \varepsilon +3)-2 \varepsilon +2) K_{\varepsilon }(a-a z)}{z} \right] .
\end{aligned}
\end{equation}

The real contribution of diagrams $a$, $b$, $e$ and $f$ as a function of $\mathbf{k}_\perp$ is
\begin{equation}
\begin{aligned} \label{eq:abefFT}
d^{\rm{a,b,e,f}}_{g \rightarrow J/\psi} (z, \mathbf{k}_\perp; \delta) & = - \frac{4^{\varepsilon-1} \pi^{2\varepsilon -1} \alpha_s^2 C_A}{(d-1)(1-\varepsilon) M^5} \frac{z^{-5} (1-z)^{-1}}{\left[\mathbf{k}_\perp^2 + \frac{M^2 (1-z)}{z^2} \right] \left[\mathbf{k}_\perp^2 + \frac{M^2 (1-z)^2}{z^2} \right]^2}\\
& \times \left[ (\mathbf{k}_\perp^2)^3 z^6 \left(-z^2+z+2 \varepsilon -2\right) \right. \\
& \left. +2 (\mathbf{k}_\perp^2)^2 M^2 (z-1)^2 z^4 \left(2 z^2 (\varepsilon -1)+z+3 (\varepsilon -1)\right) \right. \\
& \left. +\mathbf{k}_\perp^2 M^4 (z-1)^2 z^2 \left(z^4+z^3 (3-4 \varepsilon )+z^2 (2 \varepsilon -3)+z (13-12 \varepsilon )+6 (\varepsilon -1)\right) \right.  \\
& \left. +2 M^6 (z-1)^3 \left(z^3-z^2-3 z+1\right) (1- \varepsilon) \right] \, .
\end{aligned}
\end{equation}
We need the following integrals in order to obtain that contribution as a function of $\mathbf{b}_\perp$,
\begin{equation}
\begin{aligned}
I_n (z,\mathbf{b}_\perp) & = \int d^{d-2} \mathbf{k}_\perp e^{i  \mathbf{k}_\perp \cdot \mathbf{b}_\perp} \left( \frac{(\mathbf{k}_\perp^2 )^n}{\left[\mathbf{k}_\perp^2 + \frac{M^2 (1-z)}{z^2} \right] \left[\mathbf{k}_\perp^2 + \frac{M^2 (1-z)^2}{z^2} \right]^2} \right),
\end{aligned}
\end{equation}
\begin{equation}
\begin{aligned}
I_0 & = \frac{2 \pi ^{1-\varepsilon } B^{\varepsilon /2} z^{\varepsilon +2}}{M^{\varepsilon +4}} \left(\frac{K_{\varepsilon }\left(a \sqrt{1-z}\right)}{(1-z)^{\frac{\varepsilon }{2}+2}}-\frac{K_{\varepsilon }(a (1-z))-\sqrt{B} M K_{\varepsilon +1}(a (1-z))}{(1-z)^{\varepsilon +2}}\right) ,\\
I_1 & = \frac{2 \pi ^{1-\varepsilon } B^{\varepsilon /2} z^{\varepsilon }}{M^{\varepsilon+2}} \left(-\frac{\sqrt{B} M K_{\varepsilon +1}(a (1-z))}{(1-z)^{\varepsilon }}+\frac{K_{\varepsilon }(a (1-z))}{(1-z)^{\varepsilon +1}}-\frac{K_{\varepsilon }\left(a \sqrt{1-z}\right)}{(1-z)^{\frac{\varepsilon }{2}+1}}\right) ,\\
I_2 & = \frac{2 \pi ^{1-\varepsilon } B^{\varepsilon /2} z^{\varepsilon -2}}{M^\varepsilon} \left(\frac{\sqrt{B} M K_{\varepsilon +1}(a (1-z))}{(1-z)^{\varepsilon -2}}-\frac{(z+1) K_{\varepsilon }(a (1-z))}{(1-z)^{\varepsilon -1}}+\frac{K_{\varepsilon }\left(a \sqrt{1-z}\right)}{(1-z)^{\varepsilon /2}}\right) ,\\
I_3 & = \frac{2 \pi ^{1-\varepsilon } B^{\varepsilon /2} z^{\varepsilon -4}}{M^{\varepsilon-2}} \left(-\frac{\sqrt{B} M K_{\varepsilon +1}(a (1-z))}{(1-z)^{\varepsilon -4}}+\frac{(2 z+1) K_{\varepsilon }(a (1-z))}{(1-z)^{\varepsilon -3}}-\frac{K_{\varepsilon }\left(a \sqrt{1-z}\right)}{(1-z)^{\frac{\varepsilon }{2}-1}}\right),
\end{aligned}
\end{equation}
where $B= \mathbf{b}_\perp^2/4$ and $a = 2 \sqrt{B M^2/z^2}$. \\
Therefore the Fourier transform of the equation~\eqref{eq:abefFT} is the following
\begin{equation}
\begin{aligned} \label{eq:abefBT}
d^{\rm{a,b,e,f}}_{g \rightarrow J/\psi} (z, \mathbf{b}_\perp; \delta) & = - \frac{2^{2 \varepsilon -1} \pi^\varepsilon B^{\varepsilon/2} \alpha_s^2 C_A }{M^{3 + \varepsilon}  (d-1) (1-\varepsilon) } \, z^{\varepsilon} (1-z)^{-\varepsilon} \\
& \times \left[ 4 \sqrt{B} M \left(z^2-2 z+2\right) (1-\varepsilon) K_{\varepsilon +1}(a-a z) \right. \\ 
& \left. +2 \left(z^2 (1-2 \varepsilon )+2 z \varepsilon -2 \varepsilon +1\right) K_{\varepsilon }(a-a z) \right.  \\
& \left. -(z-2 \varepsilon +1) (1-z)^{\varepsilon /2} K_{\varepsilon }\left(a \sqrt{1-z}\right) \right] .
\end{aligned}
\end{equation}

\subsubsection{Behaviour at \texorpdfstring{$z \rightarrow 1$}{z1} }

In this section we make explicit the infrared divergences associated with the limit $z \rightarrow 1$ for the real diragams. We want to extract those divergences and reexpress them as poles of $\varepsilon$ and a combination of plus distributions of $z$.\\

We know that the limit form when $x \rightarrow 0$ of the modified Bessel function of second kind of order $n$, $K_n(x)$, is 
\begin{equation}
\begin{aligned}
K_0 (x) & \sim - \ln \, x \, , \\
K_{n} (x)  & \sim  \frac{\Gamma(n)}{2} \left( \frac{x}{2} \right)^{-n}  \qquad n>0 \, , 
\end{aligned}
\end{equation}
so $K_n(x)$ diverges as a logarithm when $n=0$ and as $1/x^n$ when $n>0$. We can extract this behaviour from $K_n(x)$ as follows 
\begin{equation}
\begin{aligned}
K_0(x) & =   K_0(x) + \ln \, x - \ln\, x = f(x) - \ln \, x, \\
K_n(x) & =  K_n(x) - \frac{\Gamma(n)}{2} \left( \frac{x}{2} \right)^{-n} + \frac{\Gamma(n)}{2} \left( \frac{x}{2} \right)^{-n} = g_n(x) + \frac{\Gamma(n)}{2} \left( \frac{x}{2} \right)^{-n}, \quad n>0,
\end{aligned}
\end{equation}
where $f(x)$ and $g_n(x)$ are regular at $x \rightarrow 0$.\\
On the other hand,
\begin{equation}
\begin{aligned}
    K_\varepsilon (x) & = K_0(x) + \mathcal{O}(\varepsilon^2), \\
    K_{1+\varepsilon} (x) & = K_1(x) + \varepsilon \, K_1' (x) + \mathcal{O}(\varepsilon^2). 
\end{aligned}
\end{equation}
Since the order $\mathcal{O}(\varepsilon)$ of $K_\varepsilon(x)$ is zero we can define
\begin{equation}
\begin{aligned}
    K_\varepsilon (x) & = f(x) - \ln\, x + \mathcal{O}(\varepsilon^2), \\
    K_{1+\varepsilon} (x) & = g_{1+\varepsilon}(x) + \frac{\Gamma(1+\varepsilon)}{2} \left( \frac{x}{2} \right)^{-1-\varepsilon} .
\end{aligned}
\end{equation}
with $f(x)$ and $g_{1+ \varepsilon}(x)$ regular at $x \rightarrow 0$.\\
If we proceed in the same way in eq.~\eqref{eq:cdBT} we obtain the following
\begin{equation}
\begin{aligned}
& \frac{(1-z)^{2-\varepsilon/2}}{[(1-z)^2 + \delta^2 z^2 ]} K_\varepsilon (a \sqrt{1-z})  \\
& =  (1-z) \left[ \frac{A_1(z)- (L_T - \ln(\mu^2 / M^2))/4}{(1-z)_+} - \frac{1}{2} \left( \frac{\ln(1-z}{1-z} \right)_+ + \mathcal{O}(\varepsilon^2) \right] ,
\end{aligned}
\end{equation}
\begin{equation}
\begin{aligned}
& \frac{(1-z)^{1- \varepsilon}}{[(1-z)^2 + \delta^2 z^2]} K_\varepsilon (a(1-z)) \\
& = \frac{\delta(1-z)}{2} \left( \ln \, \delta \left(L_T - \ln \frac{\mu^2}{M^2} \right) + \ln^2 \delta + \frac{\pi^2}{12}  \right)+ \frac{A_2(z) }{(1-z)_+} -  \left( \frac{\ln(1-z)}{1-z} \right)_+ + \mathcal{O}(\varepsilon) ,
\end{aligned}
\end{equation}
with
\begin{equation}
\begin{aligned}
A_1(z, b_T)  & =  K_0 \left( \frac{b_T M \sqrt{1-z}}{z} \right) + \ln \left(b_T M\frac{\sqrt{1-z} \, e^{\gamma_E}}{2} \right),\\
A_2(z, b_T) & =  K_0 \left( \frac{b_T M (1-z)}{z} \right) + \ln \left( \frac{b_T M(1-z) \, e^{\gamma_E}}{2} \right)
\,,
\end{aligned}
\end{equation}
that are regular at $z \rightarrow 1$.\\
And for the eq.~\eqref{eq:abefBT}:
\begin{equation}
\begin{aligned}
& (1-z)^{-\varepsilon} K_{1+ \varepsilon} (a(1-z)) \\
& =  B(z) + \frac{\Gamma(1 + \varepsilon)}{2} \left( \frac{\sqrt{B}M}{z} \right)^{-1-\varepsilon} \left( -\frac{\delta(1-z)}{2  \varepsilon_{\text{IR}}} + \frac{1}{(1-z)_+} - 2 \varepsilon \left( \frac{\ln(1-z)}{1-z} \right)_+ +  \mathcal{O}(\varepsilon^2) \right) ,
\end{aligned}
\end{equation}
\begin{equation}
\begin{aligned}
& (1-z)^{-\varepsilon} K_{\varepsilon}(a(1-z)) \\
& =  (1-z) \left[ \frac{A_2(z)- (L_T - \ln(\mu^2/M^2))/2}{(1-z)_+} - \left( \frac{\ln(1-z)}{1-z} \right)_+ + \mathcal{O}(\varepsilon) \right] ,
\end{aligned}
\end{equation}

\begin{equation}
\begin{aligned}
& (1-z)^{-\varepsilon/2} K_{\varepsilon}(a \sqrt{1-z}) \\
& = (1-z) \left[ \frac{A_1(z) - (L_T - \ln(\mu^2/M^2))/4}{(1-z)_+} - \frac{1}{2} \left( \frac{\ln(1-z)}{1-z} \right)_+ + \mathcal{O}(\varepsilon) \right] ,
\end{aligned}
\end{equation}
with
\begin{equation}
\begin{aligned}
B(z, b_T) & = K_{1+ \varepsilon} \left( \frac{2 \sqrt{B} M (1-z)}{z} \right) - \frac{\Gamma(1+\varepsilon)}{2} \left( \frac{\sqrt{B} M (1-z)}{z} \right)^{-1-\varepsilon} \\
& = K_1 \left( \frac{2 \sqrt{B} M (1-z)}{z}  \right) - \frac{z}{2 \sqrt{B} M (1-z)} + \mathcal{O}(\varepsilon).
\end{aligned}
\end{equation}
which is regular at $z \rightarrow 1$.\\

Note that after substituting the expansions of this section in equations \eqref{eq:cdBT} and \eqref{eq:abefBT} we obtain the final result of equations \eqref{eq:cdnorenor} and \eqref{eq:abefnorenor}.

\subsection{Integrated FF} \label{sec:FF}

In this section we show the details for the calculation of the equation \eqref{eq:FF}. 
The virtual contribution for the integrated function is the same as for the TMDFF in eq.~\eqref{eq:VirNorenor}:
\begin{equation}
\begin{aligned}
\frac{1}{z^2} d^{\text{NLO,vir.}}_{g \rightarrow J/\psi} (z ; \delta) & = \frac{\pi \alpha_s}{8 (d-1) \, m_c^3 } \frac{\alpha_s C_A}{2 \pi} \delta(1-z) \left[ \frac{1}{\veuv} \left(\frac{\beta_0}{ C_A} + 2\, \ln \delta \right) - \frac{1}{\veir }  \right. \\
 & -  \left. 2 \, \ln^2 \delta + 2\, \ln \delta \, \ln \frac{\mu^2}{M^2} +  \left( \frac{8}{3} - \frac{2 n_f}{3 C_A} \right) \ln \frac{\mu^2}{M^2} + \frac{16}{3} \ln{2} - \frac{3\pi^2}{2} + \frac{59}{9} - \frac{10 n_f}{9 C_A}  \right] .
\end{aligned}
\end{equation}
In the case of the real contribution, we start from equations \eqref{eq:cdFT} and \eqref{eq:abefFT}, and instead of computing the Fourier transform to obtain the TMDFF, we integrate over $\mathbf{k}_\perp$. 

For the contribution of the diagrams~\ref{fig:NLOreal}(c + d1 + d2) + h.c. we need the following integrals
\begin{equation}
\begin{aligned}
I_n (z) & = \int d^{d-2} \mathbf{k}_\perp \left( \frac{(\mathbf{k}_\perp^2 )^n}{\left[\mathbf{k}_\perp^2 + \frac{M^2 (1-z)}{z^2} \right] \left[\mathbf{k}_\perp^2 + \frac{M^2 (1-z)^2}{z^2} \right]} \right) \\
& =  \frac{\pi^{1-\varepsilon}}{\Gamma(1-\varepsilon)} \int_0^\infty dy \left( \frac{y^{n-\varepsilon}}{\left[y + \frac{M^2 (1-z)}{z^2} \right] \left[y + \frac{M^2 (1-z)^2}{z^2} \right]} \right)
\end{aligned}
\end{equation}
\begin{equation}
\begin{aligned} \label{Int:cdFF}
I_0 & = \frac{\pi ^{2-\varepsilon } \csc (\pi  \varepsilon ) }{z^2 \Gamma (1-\varepsilon )} \left(\frac{M^2}{z^2}\right)^{-\varepsilon -1} \left((1-z)^{-2 \varepsilon -1}-(1-z)^{-\varepsilon -1} \right) ,\\
I_1 & = \frac{\pi ^{2-\varepsilon } \csc (\pi  \varepsilon ) }{z^2 \Gamma (1-\varepsilon )} \left(\frac{M^2}{z^2}\right)^{-\varepsilon } \left( (1-z)^{-\varepsilon }-(1-z)^{1-2 \varepsilon } \right) ,\\
I_2 & = \frac{\pi ^{2-\varepsilon } \csc (\pi  \varepsilon ) }{z^2 \Gamma (1-\varepsilon )} \left(\frac{M^2}{z^2}\right)^{-\varepsilon +1} \left( (1-z)^{3-2 \varepsilon }-(1-z)^{1-\varepsilon } \right) .
\end{aligned}
\end{equation}
By using the results of the integrals shown in equation \eqref{Int:cdFF} we obtain:
\begin{equation}
\begin{aligned}
\frac{1}{z^2}
d^{\rm{c,d}}_{g \rightarrow J/\psi} (z; \delta) & = \frac{(1-z) 4^{\varepsilon -1} \pi ^{\varepsilon +1} C_A \alpha _s^2 (1-z)^{-2 \varepsilon } z^{-2 \varepsilon -1} \csc(\pi \varepsilon) \left(\frac{M^2}{z^2}\right)^{-\varepsilon }}{(d-1) M^3 (1-\varepsilon) \Gamma (1-\varepsilon ) \left( (1-z)^2+\delta ^2 z^2\right)} \\
& \times \left[  2 z^4+z^3 \left((1-z)^{\varepsilon }-6\right)  -2 z^2 \left(\varepsilon  \left((1-z)^{\varepsilon }+2\right)-5\right) \right. \\
& \left. +z \left(-(1-z)^{\varepsilon }+2 \varepsilon  \left((1-z)^{\varepsilon }+2\right)-6\right)-4 \varepsilon +4 \right]
\end{aligned}
\end{equation}
In that equation there is an infrared divergence associated with the limit $z \rightarrow 1$ which can be made explicit by using the expansion
\begin{equation}
\begin{aligned}
\frac{(1-z)^{1-2 \varepsilon}}{(1-z)^2 + \delta^2 z^2} & = \delta(1-z) \left( - \ln \, \delta + \frac{\varepsilon}{24} \left( 12 \, \ln^2 \delta + \pi^2 \right) \right) + \frac{1}{(1-z)_+} - \varepsilon \left( \frac{\ln(1-z)}{1-z} \right)_+ \\
& + \mathcal{O}(\varepsilon^2)
\end{aligned}
\end{equation}
Therefore, the final result for the contribution of the diagrams~\ref{fig:NLOreal}(c + d1 + d2) + h.c. to the integrated function FF at NLO is as follows
\begin{equation}
\begin{aligned}
\frac{1}{z^2}
d_{g \rightarrow J/\psi}^{\rm{c, d}} (z; \delta) & = \frac{\alpha_s^2 C_A}{8(d-1) m_c^3 } \left[  \frac{1}{\varepsilon} \left( -  \delta(1-z) \ln \, \delta+\frac{2 z^4-5 z^3+10 z^2-7 z+4}{4z(1-z)_+}  \right) \right. \\
& \left. + \delta(1-z) \left( \ln^2 \delta + \frac{\pi^2}{12}     \right) +  \frac{(1-z)^2  (2 z-1)}{4(1-z)_+} \right. \\
& \left. - \frac{-4 z^4+11 z^3-20 z^2+13 z-8}{4z} \left( \frac{\ln(1-z)}{1-z} \right)_+ \right. \\
& \left. + \ln \frac{\mu^2}{M^2} \left(-   \delta(1-z) \ln \, \delta +\frac{2 z^4-5 z^3+10 z^2-7 z+4}{4 z(1-z)_+} \right) \right]
\end{aligned}
\end{equation}
For the contribution of the diagrams ~\ref{fig:NLOreal}(a + b + e1 + e2 + f1 +f2) + h.c. we need the following integrals
\begin{equation}
\begin{aligned}
I_n (z) & = \int d^{d-2} \mathbf{k}_\perp \left( \frac{(\mathbf{k}_\perp^2 )^n}{\left[\mathbf{k}_\perp^2 + \frac{M^2 (1-z)}{z^2} \right] \left[\mathbf{k}_\perp^2 + \frac{M^2 (1-z)^2}{z^2} \right]^2} \right) \\
& =  \frac{\pi^{1-\varepsilon}}{\Gamma(1-\varepsilon)} \int_0^\infty dy \left( \frac{y^{n-\varepsilon}}{\left[y + \frac{M^2 (1-z)}{z^2} \right] \left[y + \frac{M^2 (1-z)^2}{z^2} \right]^2} \right)
\end{aligned}
\end{equation}
\begin{equation}
\begin{aligned} \label{Int:abefFF}
I_0 & = - \frac{\pi ^{2-\varepsilon } \csc (\pi  \varepsilon ) }{z^2 \Gamma (1-\varepsilon )} \left(\frac{M^2}{z^2}\right)^{-\varepsilon -2} \left(- (1-z)^{-\varepsilon -2}-z \varepsilon  (1-z)^{-2 \varepsilon -3}+(1-z)^{-2 \varepsilon -2} \right) ,\\
I_1 & = - \frac{\pi ^{2-\varepsilon } \csc (\pi  \varepsilon ) }{z^2 \Gamma (1-\varepsilon )} \left(\frac{M^2}{z^2}\right)^{-\varepsilon -1} \left( (1-z)^{-2 \varepsilon -1}+z \varepsilon  (1-z)^{-3 \varepsilon -1}-(1-z)^{-3 \varepsilon -1} \right) ,\\
I_2 & = - \frac{\pi ^{2-\varepsilon } \csc (\pi  \varepsilon ) }{z^2 \Gamma (1-\varepsilon )} \left(\frac{M^2}{z^2}\right)^{-\varepsilon } \left( -(1-z)^{-\varepsilon }+(z+1) (1-z)^{1-2 \varepsilon }-z \varepsilon  (1-z)^{1-2 \varepsilon } \right) ,\\
I_3 & = - \frac{\pi ^{2-\varepsilon } \csc (\pi  \varepsilon ) }{z^2 \Gamma (1-\varepsilon )} \left(\frac{M^2}{z^2}\right)^{-\varepsilon +1} \left( (1-z)^{1-\varepsilon }-(2 z+1) (1-z)^{3-2 \varepsilon }+z \varepsilon  (1-z)^{3-2 \varepsilon } \right) .
\end{aligned}
\end{equation}
By using the results of the integrals shown in equation \eqref{Int:abefFF} we obtain:
\begin{equation}
\begin{aligned} 
\frac{1}{z^2}
d^{\rm{a,b,e,f}}_{g \rightarrow J/\psi} (z)  = & \frac{\alpha_s^2 C_A 4^{\varepsilon -1} \pi ^{\varepsilon +1} (1-z)^{-3 \varepsilon -1} z^{-2 \varepsilon -3} \left(\frac{M^2}{z^2}\right)^{-\varepsilon} \csc(\pi \varepsilon)}{(d-1) M^3 (\varepsilon -1) \Gamma (1-\varepsilon)} \\
\times & \left[ z^6 \left(-4 \varepsilon ^2 (1-z)^{\varepsilon }+7 \varepsilon  (1-z)^{\varepsilon }-2 (1-z)^{\varepsilon }+\varepsilon \right) \right. \\
+ & \left. z^5 \left(4 \varepsilon ^2 \left(3 (1-z)^{\varepsilon }-1\right)+4 (1-z)^{\varepsilon }+\varepsilon  \left(2-18 (1-z)^{\varepsilon }\right)-1\right)\right.\\
+& \left.2 z^4 \left(\varepsilon ^2 \left(3-7 (1-z)^{\varepsilon }\right)+(1-z)^{\varepsilon }-(1-z)^{2 \varepsilon } \right. \right. \\
& \left. \left. +\varepsilon  \left(7 (1-z)^{\varepsilon }+(1-z)^{2 \varepsilon }-1\right)-1\right)\right. \\
+& \left. z^3 \left(14 \varepsilon ^2 \left((1-z)^{\varepsilon }-1\right)-10 (1-z)^{\varepsilon }+5 (1-z)^{2 \varepsilon }  \right.\right. \\
& \left. \left.-2 \varepsilon  \left(4 (1-z)^{\varepsilon }+2 (1-z)^{2 \varepsilon }-5\right)+6\right) \right. \\
+ & \left. z^2 \left(-18 \varepsilon ^2 \left((1-z)^{\varepsilon }-1\right)-16 \left((1-z)^{\varepsilon }-1\right)^2 \right. \right. \\
& \left. \left.+\varepsilon  \left(-9 (1-z)^{\varepsilon }+14 (1-z)^{2 \varepsilon }-5\right)\right)\right.\\
+& \left. z \left(6 \varepsilon ^2 \left((1-z)^{\varepsilon }-1\right)+19 \left((1-z)^{\varepsilon }-1\right)^2 \right. \right. \\
& \left. \left.-6 \varepsilon  \left(-5 (1-z)^{\varepsilon }+3 (1-z)^{2 \varepsilon }+2\right)\right) \right. \\\
+ & \left.6 (\varepsilon -1) \left((1-z)^{\varepsilon }-1\right)^2 
\right]
\end{aligned}
\end{equation}
In that equation there is an infrared divergence associated with the limit $z \rightarrow 1$ which can be made explicit by using the expansion
\begin{equation}
\begin{aligned}
\frac{1}{(1-z)^{1+\varepsilon}} &  = - \frac{\delta(1-z)}{\veir} + \frac{1}{(1-z)_+} - \varepsilon \left( \frac{\ln(1-z)}{1-z} \right)_+ + \mathcal{O}(\varepsilon^2) .
\end{aligned}
\end{equation}
Therefore, the final result for the contribution of diagrams~\ref{fig:NLOreal}(a + b + e1 + e2 + f1 + f2) + h.c. to the integrated function FF at NLO is as follows
\begin{equation}
\begin{aligned}
\frac{1}{z^2}
d_{g \rightarrow J/\psi}^{\rm{a,b,e,f}} (z) & = \frac{\alpha_s^2 C_A}{8(d-1)m_c^3} \left[ \frac{1}{\varepsilon} \left( \frac{\delta(1-z)}{2}-\frac{-2 z^3+3 z^2-2 z+1}{4(1-z)_+} \right) \right.\\
& \left. +   \frac{-6 z^3+13 z^2-12 z+1}{4 (1-z)_+}-\frac{4 z^5-5 z^4+4 z^3-3 z^2 }{4 z^2} \left(\frac{\ln (1-z) }{1-z}\right)_+  \right. \\
& \left. +  \ln \frac{\mu^2}{M^2} \left( \frac{\delta(1-z)}{2}-\frac{-2 z^3+3 z^2-2 z+1}{4(1-z)_+} \right)  \right] .
\end{aligned}
\end{equation}

Finally, the total real contribution to the integrated function FF at NLO is
\begin{equation}
\begin{aligned}
\frac{1}{z^2} 
d^{\rm{NLO,real}}_{g \rightarrow J/\psi} (z; \delta) & = \frac{\pi \alpha_s}{8 (d-1) \, m_c^3 } \frac{\alpha_s C_A}{2 \pi} \left[ \delta(1-z) \left( - \frac{2 \, \ln \, \delta}{\veuv} + \frac{1}{ \veir}  \right) + \frac{P_{g/g}}{ \veuv} \right. \\
& \left. + \delta(1-z) \left( -2 \, \ln \, \delta \, \ln \frac{\mu^2}{M^2} + 2\, \ln^2 \delta + \ln \frac{\mu^2}{M^2} + \frac{\pi^2}{6} \right) + P_{g/g}\, \ln \frac{\mu^2}{M^2} \right. \\
& \left. -  \frac{2(z^3 -2 z^2 +2z) }{(1-z)_+} - \frac{4(z^2 - z +1)^2}{z} \left( \frac{\ln(1-z)}{1-z} \right)_+  \right] .
\end{aligned}
\end{equation}


\bibliographystyle{utphys}
\bibliography{refs_TMDFF}

\end{document}